\RequirePackage{fix-cm}
\documentclass[natbib,smallextended]{svjour3}
\smartqed
\usepackage[english]{babel}
\usepackage{amssymb}
\usepackage{graphicx}
\usepackage{amsmath}
\usepackage{color}

\newtheorem{principle}{Principle}

\begin{document}

\journalname{}

\date{July 17th, 2015}

\title{Perfect Prediction Equilibrium}

\author{Ghislain Fourny \and St\'ephane Reiche \and Jean-Pierre Dupuy}

\institute{G. Fourny\at
	Z\"urich, Switzerland\\
	\email{ghislain.fourny@gmail.com}
	\and
	S. Reiche \at
	Mines ParisTech, France \\
	\email{stephane.reiche@mines.org}
	\and
	J.-P. Dupuy \at
	Dept. of Political Science, Stanford University, California \\
	\email{jpdupuy@stanford.edu}
}

\maketitle

\begin{abstract}

In the framework of finite games in extensive form with perfect information and
strict preferences, this paper introduces a new equilibrium concept: the Perfect
Prediction Equilibrium (PPE).

In the Nash paradigm, rational players consider that the opponent's
strategy is fixed while maximizing their payoff.
The PPE, on the other hand, models the behavior of agents
with an alternate form of rationality that involves a Stackelberg
competition with the past.

Agents with this form of rationality integrate in their reasoning that they have such accurate logical
and predictive skills, that the world is fully transparent: all players share the same knowledge and know
as much as an omniscient external observer. In particular, there is common
knowledge of the solution of the game including the reached outcome and the
thought process leading to it. The PPE is stable given each player's knowledge
of its actual outcome and uses no assumptions at unreached nodes.

This paper gives the general definition and construction
of the PPE as a fixpoint problem, proves its existence, uniqueness and
Pareto optimality, and presents two algorithms to compute it. Finally, the PPE
is put in perspective with existing literature (Newcomb's Problem,
Superrationality, Nash Equilibrium,
Subgame Perfect Equilibrium, Backward Induction Paradox, Forward Induction).


\end{abstract}


\maketitle

\section{Introduction}

In non-cooperative game theory, one of the most common equilibrium concepts is
the Subgame Perfect Equilibrium (SPE) \citep{SPE}, a refinement of the Nash
Equilibrium \citep{JNNCG}. The SPE is defined for
dynamic games with perfect information, which are mostly represented in their
extensive form.

The extensive form represents the game as a tree where
each node corresponds to the choice of a player,
and each leaf to a possible outcome of the game associated with a payoff
distribution. This embodies a Leibnizian account of rational choice in a possible-worlds system.
The SPE is obtained by backward induction, i.e. by first choosing at
the leaves (in the future) and going backward to the root.

While the Nash equilibrium is widely accepted by the game theory community, there is also
a large consensus that it does not account for all real-life scenarios. This is because
people do not always act rationally in situations where their emotions impact their decisions,
but also because different people may have different forms of rationality, Nash describing one of them.
The Perfect Prediction Equilibrium (PPE), presented in this paper, is based on a different
form of rationality and accounts for real-life situations that Nash does not predict,
such as asynchronous exchange or promise keeping.

\subsection{Two forms of rationality}

A crucial assumption made by the Nash equilibrium in general and the SPE in particular is
that each player considers the other
player's strategy to be independent of their own strategy. Concretely, this means that the
other player's strategy is held for fixed while optimizing one's strategy. In the extensive form, it means that, at each
node, the player whose turn it is to play considers that former moves (especially the other
player's former moves) and her own current choice are independent of each other. Consequently,
the past can be taken out of the picture and only the remaining subtree is used to decide on
the current move. This motivates Backward Induction reasoning. 

This line of reasoning is intuitive to many, because the past is causally independent of the future
--- which is supported by the fact that nothing
has been observed in nature so far that would contradict it.

However, concluding that a rational player \emph{must} make this assumption
would be fallacious, as this relies on a confusion
commonly made between causal dependency on one side, and counterfactual dependency
on the other side. There are concrete examples of situations where two causally independent
events are counterfactually dependent on each other. This is often referred to as a statistical
dependency in physics (consider quantum measurements done on two entangled photons),
and which is of a nature fundamentally different than that of causality.

The relationship between a move and its anticipation has been at the core of numerous discussions in game theory. There is an apparent conflict between a player's freedom to make any choice on the one hand, and the other player's skills at anticipating the moves of his opponent on the other hand. 

The Nash equilibrium addresses this tension with the approach described a few paragraphs above. Though, having a different model on the relationship between the past moves and the current
move can lead to a different form of rationality that is no less meaningful than the Nash equilibrium approach. A rational
player could consider that her decisions are transparent to, and anticipated by former players, and, hence,
that she might want to integrate into her reasoning the past's reaction to her anticipated
current move. Concretely,
this means that a player reasoning that way would think of her relationship with the past
not through a Cournot-like competition (that would be the Nash reasoning), but rather through
a Stackelberg-like competition (that would be the reasoning in this paper) \footnote{The analogy with Cournot vs. Stackelberg is used to help understand the paradigm shift underlying the PPE. Classically, Stackelberg competition is built on the future's reaction function. In the PPE, we mean Stackelberg competition built on the past's reaction function, which means the reaction function of the past to the \emph{anticipation} of a move, embedding excellent anticipation skills of the opponent, as opposed to a frozen past in the Nash paradigm.}.

Newcomb's problem (see Section \ref{section-newcomb}) illustrates that both approaches (Cour\-not-like, Stackel\-berg-like) are commonly
taken by people in the real world, and that people can feel very strong taking one side or the other.
This problem sets up a very simple situation with an action
in the past (predicting and filling boxes) and a move in the present (picking one or two boxes), and on
purpose leaves open how they depend on each other. Newcomb's problem
demonstrates that some people (the \emph{two-boxers}) make the assumption that
past actions are counterfactually independent of their own moves, which corresponds
to what could be called two-boxer rationality (Cournot competition with the predictor).
Maybe this explains why most game theoreticians are in this category.
However, many other people (the \emph{one-boxers})
consider that past actions and their own moves are counterfactually dependent -- entangled.
Making this assumption is no less rational, as their line of reasoning then comes down
to optimizing their utility as well, with a Stackelberg view of the relationship with the predictor's action.

This Stackelberg competition with the past by no means takes away the agents' freedom to make
decisions, nor does it break the laws of physics: the dependency with the past is purely
based on counterfactual reasonings such as "If I were to make this move, the other
player would have predicted it and acted in this other way." There is no such thing
as a causal impact on the past or as changing the past.

Let us call this alternate form of rationality one-boxer rationality \footnote{\label{footnote-rationality}Asking an agent how many
boxes he would pick in Newcomb's problem is one of the most convenient ways to tell the difference
between the two kinds of rationalities. Hence, we use the terms ``one-boxer rational'' for agents that would
pick one box, and that would reach the PPE in an extensive-form game, and ``two-boxer rational'' for
agents that would pick both boxes, and would reach the SPE in an extensive-form game. Other possibilities would be
Nash rational vs. fixpoint rational, projected time rational vs. occuring time rational, etc. }.

Most equilibria in game theory are refinements of the Nash equilibrium. They account for
the behavior of players that are two-boxer-rational, while little work has been done for
one-boxer-rational players. This paper introduces the solution concept corresponding
to the equilibrium reached by one-boxer-rational players: the Perfect Prediction Equilibrium (PPE).

\subsection{The assumptions}

This paper defines a solution concept for games that are:
\begin{description}
\item [in extensive form,] i.e., the game is represented by an explicit tree, rather than by
a flat set of strategies for each player; The game structure is Common Knowledge.
\item [with perfect information,] meaning that for each choice they make, players know
at which node they are in the tree;
\item [with strict preferences\footnote{Other formulations found in literature include ``in
general position'' \citep{RABICKR}, ``nondegenerate'' \citep{Halpern2001} },] meaning
that given two outcomes, players are never indifferent between those;
\item [played by rational players,] in its widely accepted meaning in game theory, i.e., they
are logically smart and play to the best of their interests, preferences and knowledge
with the goal of maximizing their utility; The
player's rationality is Common Knowledge.
\item [played by agents with a one-boxer form of rationality] -- that is, with a Stackelberg
account of the past's reaction, which is the novelty of this paper. The one-boxer rationality
of the players is Common Knowledge as well.
\end{description}

We are interested in games exclusively played by one-boxer-rational agents. Past moves are
considered to be counterfactually dependent on the current player's move: Would a player play differently than he actually does,
then it would be the rational thing to do, and the other player would have anticipated it.
This leads to a form of Stackelberg competition where the current player considers the past's reaction function.
The reaction function can be
seen as literally reading each other's minds (or at least, as believing that the other player
does so, which is sufficient). These players consider that there is Common
Knowledge (CK) of the solution of the game: CK of the outcome of the
game and of each other's thought processes. Hence, one-boxer rationality can also be seen
as the belief that the world is totally transparent, and that
the players have as much knowledge as an omniscient external observer (Perfect Prediction).

\subsection{Outline of the paper}

In the remainder of this paper, and for the sake of a smooth read, we will consider,
like the one-boxer-rational players,
that it holds that the world is totally transparent and that there is CK
of the solution (i.e., of all actual moves) of the game, embracing their account of the world.

Our starting point is the question: \emph{if there were an
equilibrium which is totally transparent to itself, what would it be?}

Starting with the Perfect Prediction abilities (assumed
by the players), two principles are postulated (1. Preemption, 2. Rational Choice).

Using these two
principles, we can show that the equilibrium reached by one-boxer-rational players, and which
we call Perfect Prediction Equilibrium (PPE), exists and is unique. Hence, this answers the question asked
above: \emph{assuming that there is an equilibrium totally transparent to
itself, then it is necessarily the Perfect Prediction Equilibrium}. Furthermore,
because of its uniqueness, the players, aware of the assumptions, are able to
unambiguously calculate the PPE and react accordingly, which leads them to this
very equilibrium and makes their prediction correct (self-fulfilling prophecy).
This closes the loop: the PPE is totally transparent to itself.

In addition, the Perfect Prediction Equilibrium happens to have the
property of Pareto-optimality: there is no other outcome giving a better payoff
to both players.

Hence, all conjectures (existence, uniqueness, Pareto-optimality) made by
\citet{JPDPFNCE} are proven in this paper for the PPE.

In Section \ref{section-perfect-prediction}, we introduce Newcomb's problem as an illustration
of the importance of assuming (or not) counterfactual independence. We then explain the relationship between total transparency and Perfect Prediction. We posit
two principles, which we apply to some examples. In Section \ref{section-construction}, we give the
general definition of the PPE as well as its construction, and prove its
existence, uniqueness and Pareto optimality. Finally, Section \ref{section-related-work} gives links to
existing literature on Superrationality, the Backward Induction Paradox, and Forward Induction\footnote{In our paper, we use the
expression ``forward induction'' with two meanings. The first one is forward
induction by construction: reasoning starts at the root and finishes at an
outcome. The second one is the meaning that Forward Induction has accumulated in
the history of game theory and which is explained in Section \ref{section-related-work}. For clarity,
we refer to the former without capitals ``forward induction'', and to the latter
with capitals ``Forward Induction''.}. In the annex, we provide more technical background for readers
interested in deeper details, such as the analysis of the underlying preemption structure of the PPE, from which two algorithms can be derived to compute it. The annex also gives complementary material such as proofs of the lemmas and theorems, a complete analysis of biped games, and the equations behind the PPE.

\section{Perfect Prediction}

Before giving the formal definition of the PPE and proving its properties, we focus on a few examples, explain the general semantics
of Perfect Prediction and of the PPE, and show how it applies to these examples.

\label{section-perfect-prediction}

\subsection{Examples used throughout this paper}

There are three games used repeatedly as examples in this paper.

\subsubsection{Take-or-Leave game}

\begin{figure}[htbp]
\centering
\includegraphics[width=8.4cm]{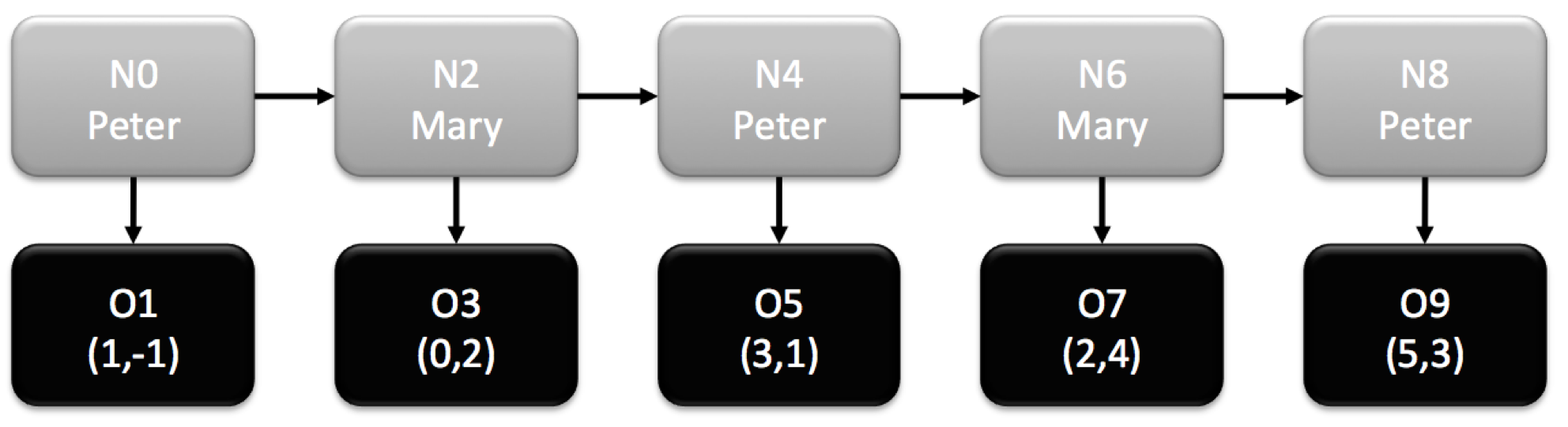}
\caption{The Take-Or-Leave game (n is odd)}
\label{Fig_TOL}
\end{figure}

The Take-or-Leave game is represented on Fig. \ref{Fig_TOL} and goes as follows:
``A referee places dollars on a table one by one. He has only $n$ dollars. There
are two
players who play in turn. Each player can either take the dollars that have
accumulated so far,
thereby ending the game, or leave the pot on the table, in which case the
referee adds one dollar to
it, and it's the other player's turn to move.''

The original TOL game does not satisfy the strict preference assumption\footnote{However,
once familiar with the PPE, the reader may notice that this assumption can sometimes
be relaxed, and that the original TOL game does have a PPE, too},
so that this example was slightly modified in such a way that the loser of
the game still gets increasing payoffs. This does not modify its (inefficient
as we will see) Subgame Perfect Equilibrium.

This type of game has often been used to describe the Backwards Induction Paradox in literature (see Section \ref{section-backward-induction-paradox}).

\subsubsection{Assurance game}

\begin{figure}[htbp]
\centering
\includegraphics[width=8.4cm]{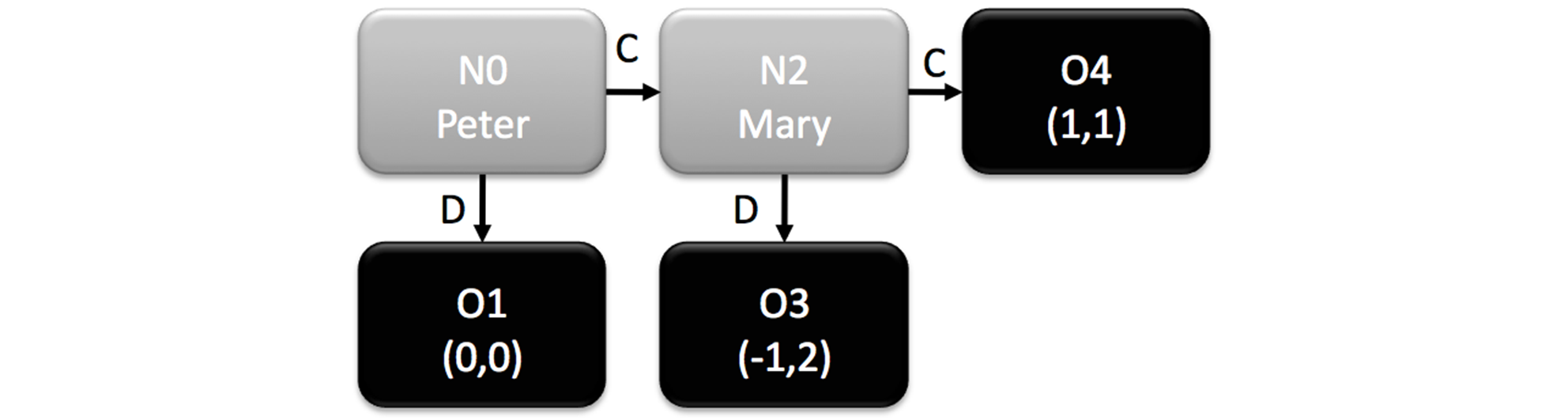}
\caption{The assurance game}
\label{Fig_Promise}
\end{figure}

The assurance game (see Fig. \ref{Fig_Promise}) models a basic, asynchronous,
Pareto-im\-proving exchange in economics (payoffs are utilities). The rule is as
follows. First, Peter has the choice between deviating
(D - both players get $0$) and cooperating (C). If Peter cooperates, then
Mary has the choice between cooperating as well (C), in which case both
players get $1$, and deviating (D), in which case Mary gets $2$ and
Peter gets $-1$.

This can be interpreted as the possibility for Peter to trust Mary or not, and if he does trust her, Mary can
choose to cooperate or not.

\subsubsection{$\Gamma$-game}

\begin{figure}[htbp]
\begin{center}
\includegraphics[width=8.4cm]{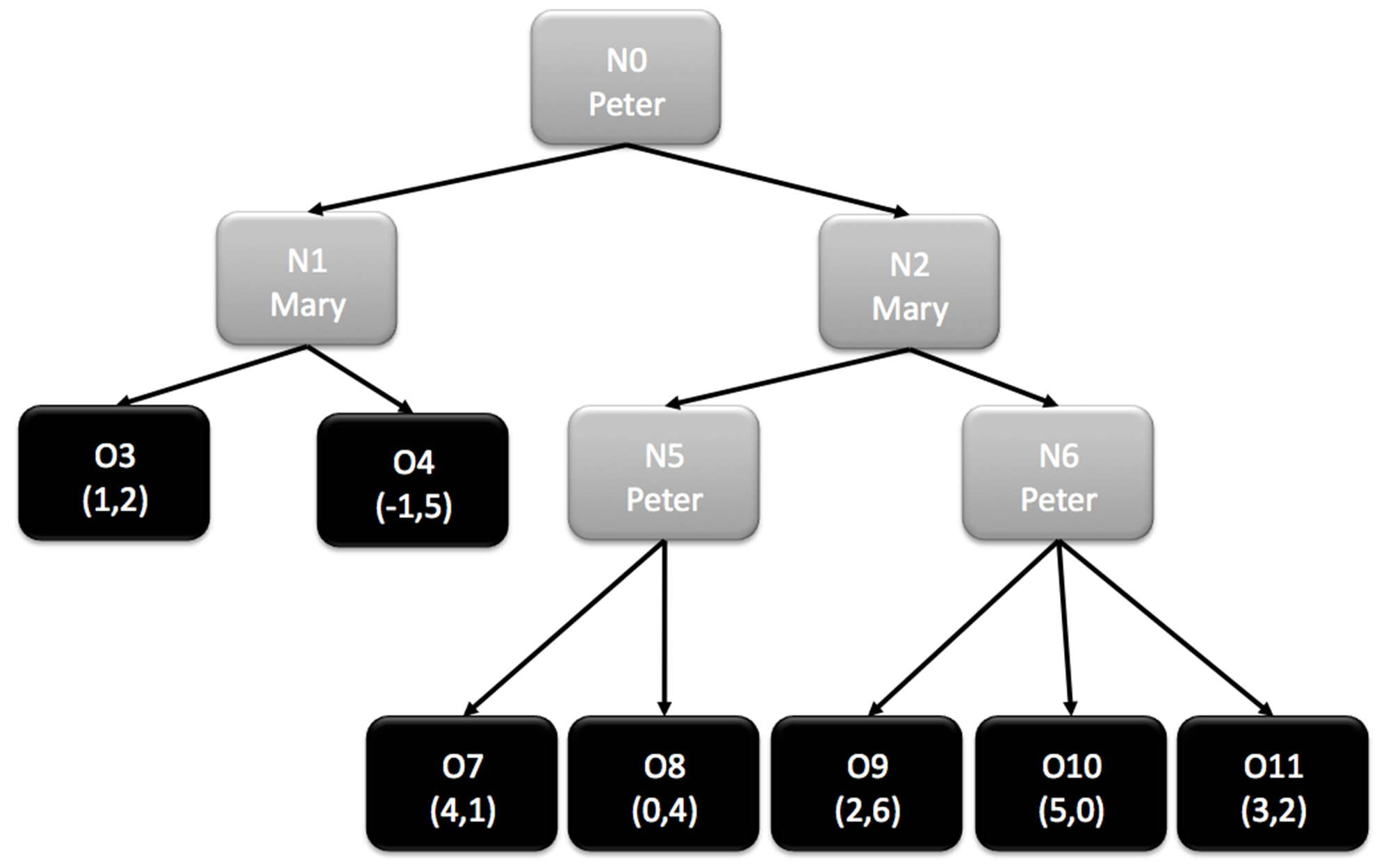}
\end{center}
\caption{The $\Gamma$-Game}
\label{Fig_Gamma}
\end{figure}

The $\Gamma$-game (see Fig. \ref{Fig_Gamma}) aims at demonstrating how the PPE is
computed on a slightly more complex game. First, Peter has the choice between
two nodes. If he chooses $n_1$, then Mary chooses between two outcomes; if he
chooses $n_2$, Mary chooses between two nodes and it is Peter's turn again. This
game does not have a particular interpretation - it was simply designed for the
reasoning to be interesting and to help the reader understand how the Perfect
Prediction Equilibrium works on general trees.

\subsection{Newcomb's Problem}
\label{section-newcomb}
Newcomb's problem \citep{NP} is an illustration of how the concepts of prediction and decision
interact, and a good starting point to introduce Perfect Prediction. In particular, it is a very good
test for distinguishing between agents with two kinds of rationalities that differ by the correlation,
or absence thereof, between prediction and decision. Historically, Newcomb's
problem, through the research and literature around it, is actually at the origin of the discovery
of these two kinds of rationalities, which is why, in this paper, we decided to call them
one-boxer rationality and two-boxer rationality, the latter being the one commonly accepted as
game-theoretical perfect rationality. The use of these names, however, by no means
reduces the scope of these rationalities to Newcomb's problem and alternate names are suggested
in footnote \ref{footnote-rationality}.

Imagine two boxes. One, B, is transparent and contains a thousand dollars; the
other, A, is opaque and contains
either a million dollars or nothing at all. The choice of the agent is either
C1: to take only what is in the opaque box, or
C2: to take what is in both boxes. At the time that the agent is presented with
this problem, a Predictor has already
placed a million dollars in the opaque box if and only if he foresaw that the
agent would choose C1. The agent knows
all this, and he has very high confidence in the predictive powers of the
Predictor. What should he do?

A first line of reasoning leads to the conclusion that the agent should choose
C1. The
Predictor will have foreseen it and the agent will have a million dollars. If he
chose C2, the opaque box would have been empty and he would
only have a thousand. The paradox is that a second line of reasoning (taken by most
game theorists as it is consistent with Nash) appears to
lead just as surely
to the opposite conclusion. When the agent makes his choice, there is or there
is not a million
dollars in the opaque box: by taking both boxes, he will obviously get a
thousand dollars more in
either case. This second line of reasoning applies dominance reasoning to the
problem, whereas the
first line applies to it the principle of maximization of expected utility.

The paradox can be solved as follows (although, to be fair, there is
to date no established consensus about it). Two-boxers assume
that the choice and the prediction are \emph{counterfactually} independent of
each other. In other words, once the prediction is made, the choice is made
independently and the prediction, which lies in the past, is held for fixed. Even
if the prediction is correct (the player gets \$ 1,000), it \emph{could have been made wrong}
(she would have gotten \$ 0).
One-boxers assume on the contrary that the prediction and the choice
are \emph{counterfactually} dependent. In other words, the prediction is correct (player gets
\$ 1,000,000),
and if the player had made the other choice, the predictor would have correctly
anticipated this other choice as well (she would have gotten \$ 1,000).

One-boxers and two-boxers are both rational: they merely make different, not to say opposite,
assumptions. This illustrates that there are at least two ways of being rational.

A probabilistic illustration along the same lines as this solution
has recently been given by \citet{Wolpert2011}. Technically, counterfactual dependence means
statistical dependence if events and decisions are seen as random variables.
It is a symmetrical relation and ignores the arrow of time
(unlike causal dependence).

The Subgame Perfect Equilibrium makes the (``two-boxer'') assumption that decisions are counterfactually independent of the past. More generally, Nash Equilibria make the (``two-boxer'') assumption that the other players' strategies are held fixed while a single player optimizes her strategy. Perfect Prediction Equilibria, on the other hand, relies on the ``one-boxer'' assumption of counterfactual dependency. In both cases, players are rational and maximize their utility to the best of their logical abilities. They
are, however, rational in two different ways.

\subsection{Transparency and Perfect Prediction}

In classical game theory (Nash equilibrium), two-boxer-rational agents know that they are rational, know that they know, etc. This is
called common knowledge (CK) of rationality in literature. Common knowledge is a form of transparency. In the case of SPE, the possible futures are each considered by, and are transparent to, the current player, for him to decide on his next move.

However, this transparency does not hold the other way round. The player being \emph{simulated} or \emph{anticipated} in a possible future -- first and foremost, in a subgame that is not actually on the path to the SPE -- cannot be fully aware of the rest of the game (that is, of the overall SPE), as this knowledge would be in conflict with the fact of being playing in the subgame at hand. The past is opaque to her, meaning that she cannot be fully aware of the past thought processes. She simply assumes that the past leading to her subgame is fixed, and the other player knows that she does (CK of \emph{two-boxer-}rationality). This is one-way transparency. Section \ref{section-backward-induction-paradox} goes into more details on what is known as the backward induction paradox.

It goes otherwise for one-boxer-rational agents. In a game played by one-boxer-rational agents, there is common knowledge of rationality as well, in that players know (and know that they know, etc) that they both react to their knowledge in their best interest. But there is more importantly CK of \emph{one-boxer-}rationality, which means that (i) each player considers himself to be transparent to the \emph{past}, in that they integrate this very CK of one-boxer-rationality in their reasoning and consider themselves to be predictable, and (ii) each player P considers himself to be transparent to the \emph{future} as well, in that they know that a future player F considers player P's moves to correlate (counterfactually) with their (F's) decision, whichever way F decides.

This more stringent form of transparency drastically changes the reasoning: since transparency goes both ways, the entire solution of the game (all thought processes, the equilibrium, etc) is CK. In such a perfectly transparent world, each agent would have the same knowledge of the world as an external omniscient spectator, this fact being CK among the agents \footnote{Some might suggest that this could be considered a Principle Zero, in addition to the two principles described in Section \ref{section-two-principles}. Since we consider this to be part of the fundamentals of the PPE framework though, as opposed to the two principles, which are more algorithmic, we decided not to call it so.}. The
players (profactually) predict the solution correctly, and would also have (counterfactually) predicted it correctly if it had been different. This is what we call Perfect Prediction. This different relation to time was called Projected Time (and the widespread relation to time Occurring Time) by \citet{JPDPFNCE}.

Rather than considering all possible futures, the one-boxer-rational agents assume that there is one solution of the game that is CK, and based on this assumption, compute and find that very solution of the game that is CK. The players react to their
CK of the outcome of the game, play accordingly and reach an outcome which is indeed the predicted outcome. This is nothing else than a fixpoint problem, which can be solved and which has a unique solution, as demonstrated in this paper.

A side effect of assuming CK of the game solution is that the construction of the equilibrium in this paper is supported by reasoning only on the equilibrium path. There is one timeline, and it is completely transparent to itself, by its definition as a fixpoint.

\subsection{Two principles behind the computation of the Perfect Prediction
Equilibrium}
\label{section-two-principles}

A Perfect
Prediction Equilibrium is a path on the tree which can be reached under the
assumption of total transparency, or, in other words, by one-boxer rational players
that genuinely believe that there is total transparency.

The computation of PPEs by the players or an external game theorist, which is
the same, is done by eliminating outcomes of the game which lead to a
contradiction (Grandfather Paradox\footnote{The Grandfather Paradox is as
follows: a time traveler goes back in time and provokes the death of his
grandfather, preventing him from meeting his grandmother. Hence, the time traveler was not born
and cannot have traveled back in time. While one might see this as a proof that
time travel is impossible, another way to look at it is that the actual timeline
is the solution of a fixpoint equation: the past, tampered with by the time
traveler, must cause the time traveler's birth for him to be able to go back
in time. In terms of game theory, we also encounter a Grandfather paradox in the
following case: an outcome is commonly known to be the outcome of the game;
having predicted it, a player deviates from the path leading to it, making the
outcome out of reach. Just like for the time traveler, we are looking for the
timeline which is immune to the Grandfather paradox: the outcome which is
actually caused by its prediction.}), i.e., which cannot possibly be commonly known as
the equilibrium.

At each step of the reasoning, there is CK of the outcomes which
are logically impossible (this is part of the thought process). Using this knowledge, it is possible to discard
additional outcomes which cannot be commonly known as the equilibrium reached by
the game, thanks to the two principles.

\subsubsection{First principle}

The CK of the outcomes which have been proven logically impossible
as well as the assumed CK of the outcome $o$ reached by the game
give the players a power of preemption; a player will deviate from the path
connecting the root to $o$ (we call this path \emph{causal bridge})
if the deviation guarantees to him a better payoff than $o$
(whatever the continuation of the game after the deviation can be, and taking into
account that some outcomes have been proven logically impossible).

\begin{figure}[htbp]
\begin{center}
\includegraphics[width=8.4cm]{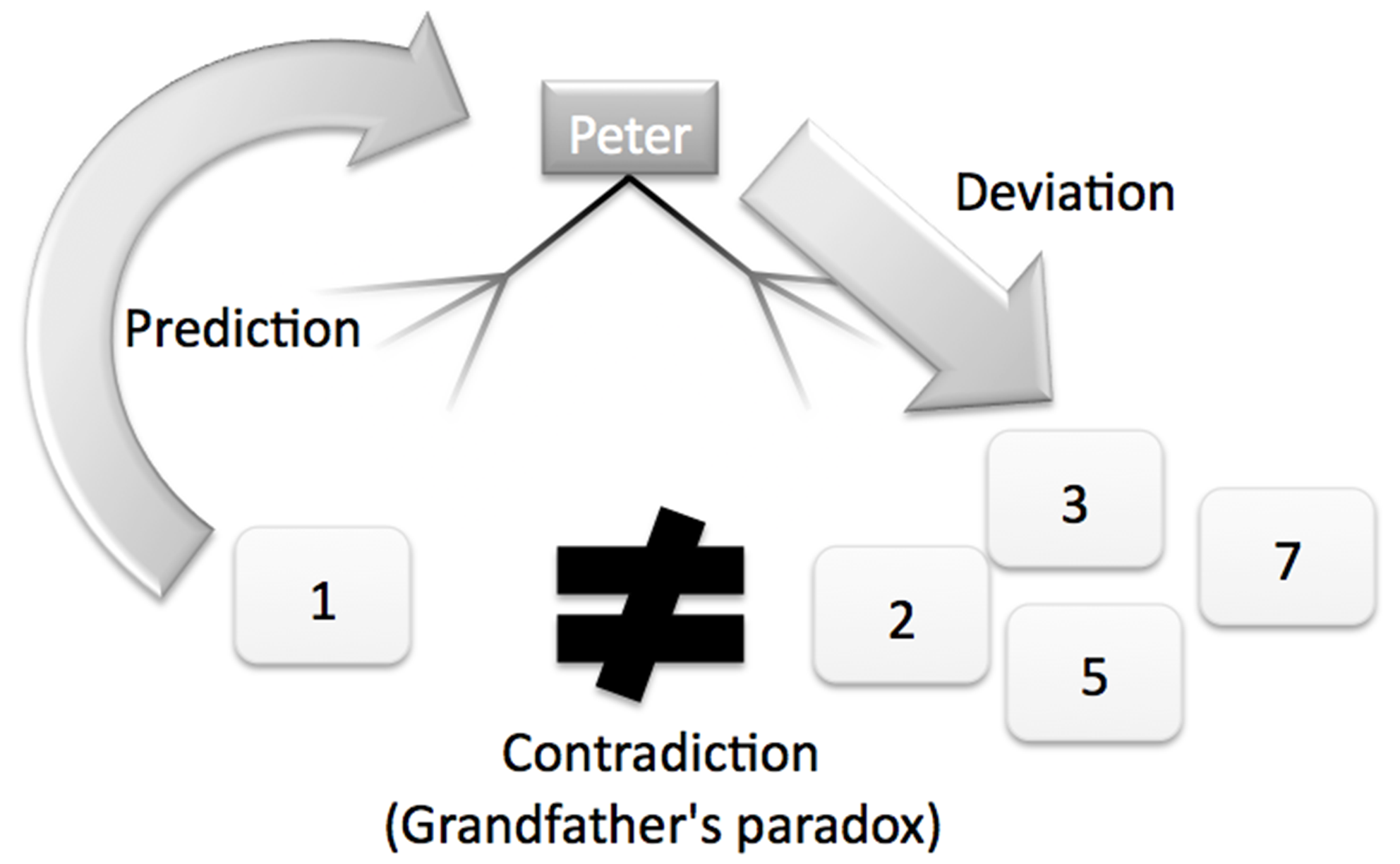}
\end{center}
\caption{Illustration of the First Principle}
\label{Fig_Principle1}
\end{figure}

If such is the case, then $o$, which is not a fixpoint, cannot be commonly known
as the solution and is logically impossible; it is said to be preempted \footnote{With
the terminology of the introduction, the one-boxer-rational agent forbids himself to pick $o$
because it would lead to an inconsistency in his reasoning pattern.}. We call
preemption this process of breaking the causal bridge
in reaction to the anticipation of an outcome (see Fig. \ref{Fig_Principle1}).

\begin{example} Let us take $o_3$ in the assurance game (Fig. \ref{Fig_Promise}).
There are no outcomes proven impossible yet. We assume that $o_3$ is the outcome
eventually reached by the game, which is known to the players. For Peter playing
at the root, a deviation from $n_2$ (which is on the path leading to $o_3$) to
$o_1$ guarantees him a better payoff (0) than what he gets with $o_3$ (-1). This
means that anticipating $o_3$, Peter deviates and breaks the causal bridge
leading to $o_3$. Hence, $o_3$ cannot be the solution: it is not possible for
Peter to predict $o_3$ \emph{and} to play towards $o_3$. $o_3$ is preempted and
is logically impossible. It can be ignored from now on.
\end{example}

\begin{example}
In the $\Gamma$-game example, $o_4$ can also be discarded this way: For Peter
playing at the root, a deviation from $n_1$ (which is on the path leading to
$o_4$) guarantees him a better payoff (4, 0, 2, 5 or 3) than what he gets with
$o_4$ (-1). $o_4$ is preempted and can also be ignored from now on.
\end{example}

\begin{principle} Let $I$ be the set of the outcomes that have not been
discarded yet. The discarded outcomes are ignored. Let $n$ be a given node which
is reached by the game and where player $p$ is playing. Let $o \in I$ be an
outcome among the descendants of $n$. $o$ is said to be preempted if all
outcomes in one of the non-empty \footnote{i.e., a subtree that only has
discarded outcomes cannot be considered} subtrees at $n$, give to $p$ a greater
payoff than $o$. A preempted outcome is logically impossible, is hence discarded
and must be ignored in the remainder of the reasoning.
\end{principle}

\subsubsection{Second principle}

\begin{figure}[htbp]
\begin{center}
\includegraphics[width=8.4cm]{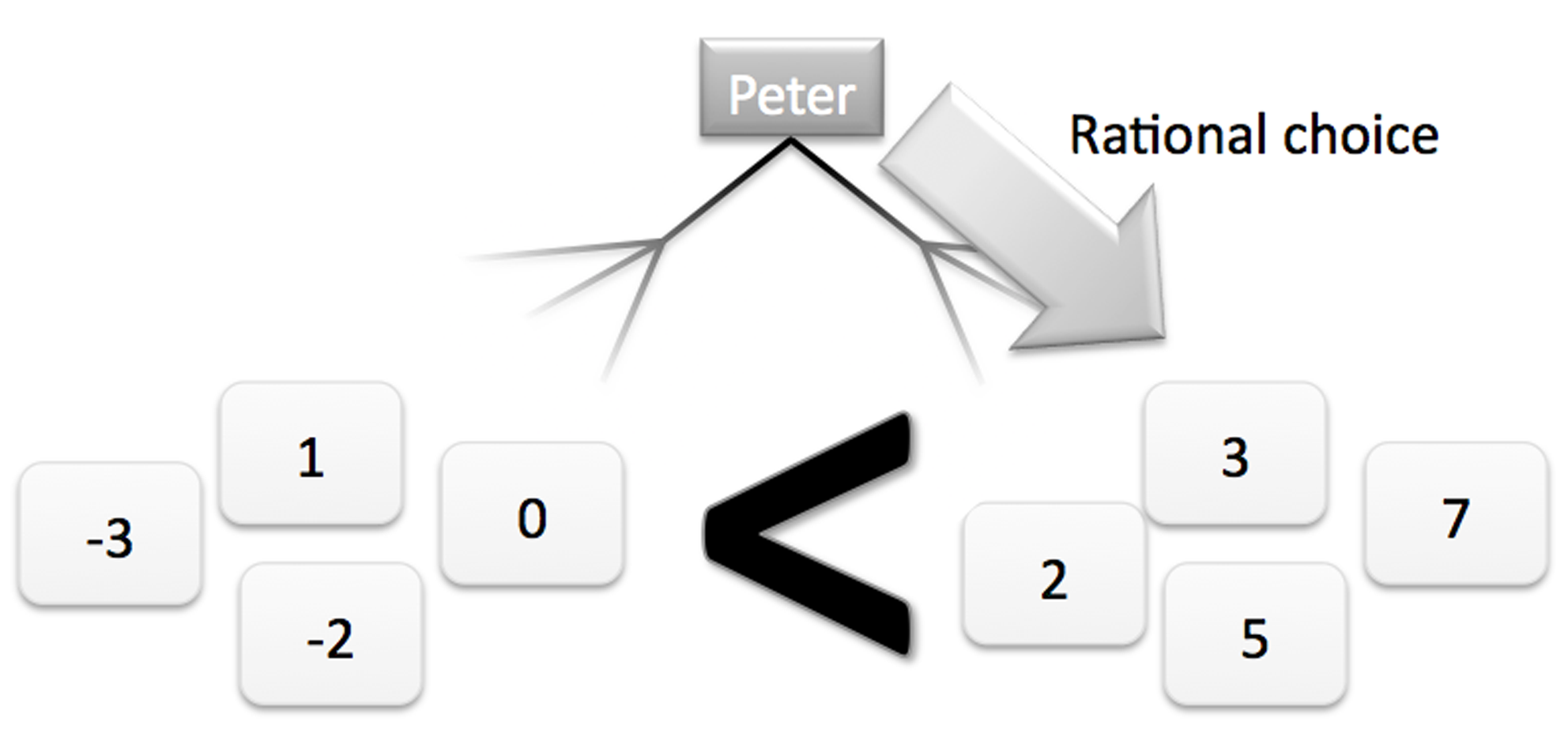}
\end{center}
\caption{Illustration of the Second Principle}\label{Fig_Principle2}
\end{figure}

In the assurance game (Fig. \ref{Fig_Promise}), given the knowledge that $o_3$ is
logically impossible, Peter is faced with the choice between $o_1$, which gives
him 0, and $n_2$, which gives him eventually for sure 1 since $o_3$ cannot be
chosen. Because he is rational, he chooses $n_2$. This is the second principle
(see Fig. \ref{Fig_Principle2}).

\begin{principle} Let $I$ be the set of the outcomes that have not been
discarded yet. The discarded outcomes are ignored. Let $n$ be a given node which
is reached by the game and where player $p$ is playing. If all outcomes in one
of the non-empty subtrees at $n$, of root $r$, are better than any other outcome
in the other subtrees, then $p$ chooses $r$. As a consequence, $r$ is also
reached by the game, and all remaining outcomes in other subtrees than the one
at $r$ are discarded.
\end{principle}

The second principle is only applicable if all outcomes on the one side are
better than all outcomes on the other side. If the payoffs are ``interlaced'',
it does not apply - but in this case, the first principle can be used instead.

Also, the second principle applies in the degenerate case where there is only
one unique non-empty subtree. Then the player chooses this subtree.

\subsubsection{The reasoning to the equilibrium}

Based on the two principles, it is possible to build a system of first-order-logic
equations for each game, as described in the annex (Section \ref{section-equations}).
By definition, a PPE is a solution to this system.
The equations are divided into three groups:
\begin{enumerate}
\item Application of the First Principle
\item Application of the Second Principle
\item Causal bridge enforcement (i.e. an equilibrium must be a consistent path starting
from the root)
\end{enumerate}

This system can be solved with a technique
which turns out to be a forward induction: one starts at the root, discards all
outcomes preempted at this step with the first principle, then moves on to the
next node - there is always exactly one - with the second principle (this node is now
known to be reached by the game),
discards all outcomes preempted at this step with the first principle,
moves on to the next node with the second principle, etc., until an outcome is
reached.

In the reasoning, the list of logically impossible outcomes is initially empty
and grows until no more outcomes can be discarded. In fact, as will be proven in
Section \ref{section-construction}, under the assumption of strict preferences, exactly one outcome
remains. The one-boxer rational players have no interest in deviating from the path leading to it
and will play towards it: it is the Perfect Prediction Equilibrium, which
always exists and is unique.

This system of equations can be described formally,
however for pedagogical reasons, from now on, we will reason directly by performing
outcome eliminations.

The reasoning is totally transparent and can be performed, ahead of the game or
without even playing, by a game theorist or any player, always leading to the
same result. Actually, as the structure of the reasoning is a forward induction,
it would even be possible to perform the reasoning \emph{as you go} during the game.

\subsection{Computation of the Perfect Prediction Equilibrium for the three
examples}
\label{applicationtoexamples}
We now show how the forward induction reasoning described above can be applied
to the examples.

\subsubsection{Assurance game (Fig. \ref{Fig_Promise})}

We begin at the root.

Being Perfect Predictors, the players at the root cannot predict $o_3$ as the
final outcome because of the first principle: the prediction of $o_3$ would lead
Peter to deviate at the root towards $o_1$, which invalidates this very
prediction. We say that outcome $o_3$ has been preempted by the $o_1$ move.
Therefore, outcome $o_3$ has been identified as not being part of the solution:
it cannot be commonly known as the outcome reached by the equilibrium and is
ignored from now on.

This can also be formulated as a reasoning in Mary's mind. She is one-boxer rational.
Mary includes Peter's past move as a reaction function to her choice between $o_3$
and $o_4$. Were she to pick $o_3$, Peter would have anticipated it as the rational
thing for her to do -- and he would have picked $o_1$. But this would be inconsistent
with her playing at $n_2$. Since picking $o_3$ is correlated with a past reaction (to the
anticipation of $o_3$) that is incompatible with $o_3$ (Grandfather's Paradox),
she cannot (one-boxer)-rationally pick $o_3$.

Now that $o_3$ is (identified as being) impossible, we are left with $o_1$ and
$o_4$, and we can apply the second principle: Peter knows that Mary is one-boxer
rational and will not pick $o_3$. He can either move
to $o_1$ and win 0, or move to $n_2$ and win 1. He chooses the
latter\footnote{Some might see a paradox here, since outcome $o_1$ has
contributed to the preemption of $o_3$, but at the same time is impossible
because not reached. Actually, there is no paradox. A proposition (``outcome
$o_3$ is part of the solution'') which implies a contradiction (``Peter
deviates, so $o_3$ is not part of the solution'') is false, but this does not
imply that a proposition (``Peter deviates'') implied by this proposition is
logically true, since false propositions logically imply anything, true or false.
In the formalization of the PPE as solution of a
first-order-logic equation system, this appears clearly.
}
and $n_2$ is known to be reached by the game.

At $n_2$, Mary, according to a trivial application of the second principle (she
can only choose $o_4$), moves to $o_4$, which is known to be reached by the
game: it is the only remaining outcome. If there is an equilibrium, it is $o_4$.

\begin{figure}[htbp]
\begin{center}
\includegraphics[width=8.4cm]{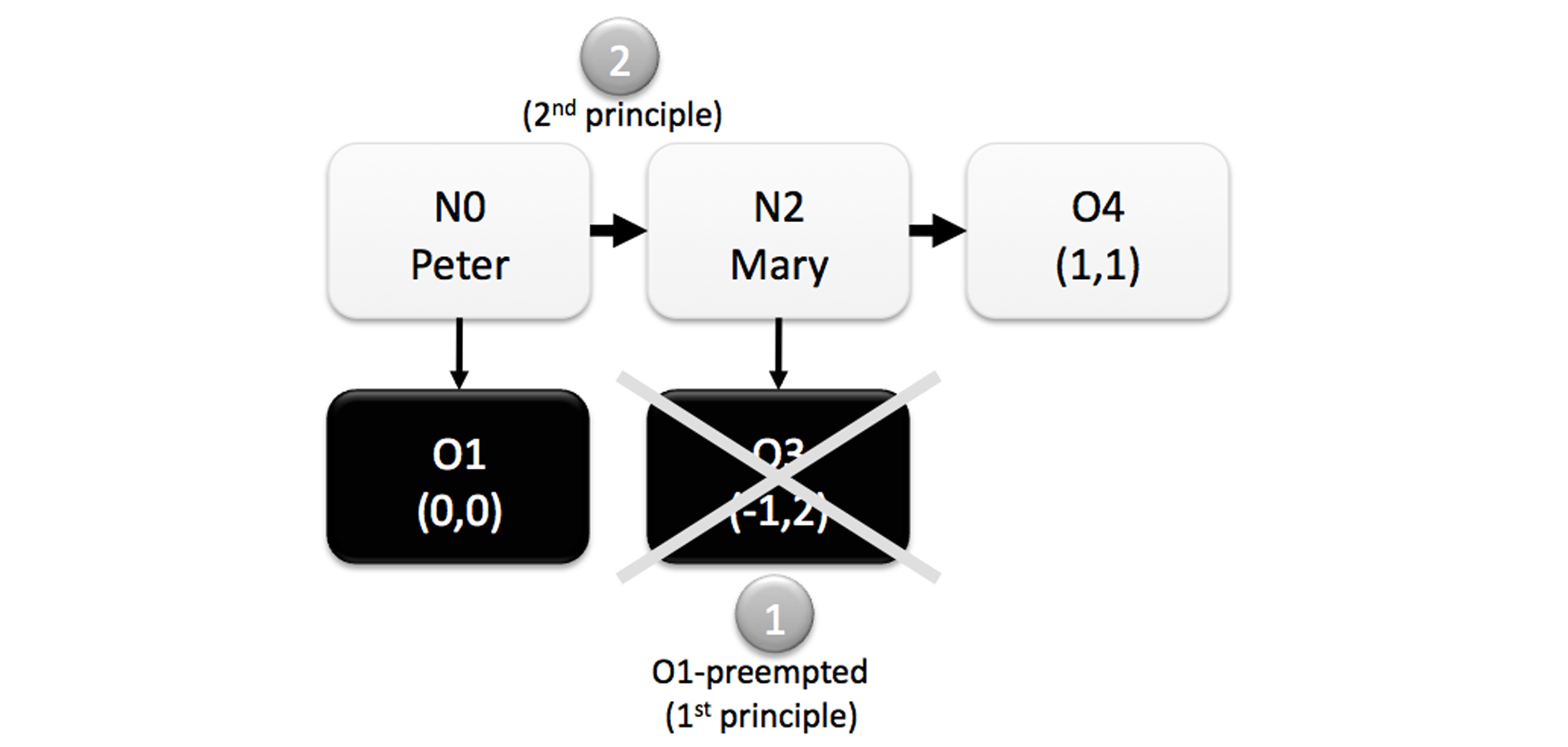}
\end{center}
\caption{The Perfect Prediction Equilibrium in the Assurance
Game}\label{Fig_Promise_Solved}
\end{figure}

Conversely, it is possible to check that this outcome corresponds to an
equilibrium. Predicting $o_4$, Peter will not deviate: according to the second
principle, he moves to $n_2$, i.e., towards $o_4$. Mary will not deviate either,
since she cannot choose $o_3$, which is preempted. More precisely, she forbids to herself
to pick $o_3$ because her reasoning as a one-boxer rational player stamped it
as an illogical move. Like a self-fulfilling
prophecy, $o_4$ is reached by the game: it is the equilibrium (Fig.
\ref{Fig_Promise_Solved}).

\subsubsection{Take-Or-Leave game (Fig. \ref{Fig_TOL})}

The previous reasoning can easily be generalized to the
Take-Or-Leave-game (Fig. \ref{Fig_TOL}).

One begins at the root. The application of the first principle here leads to the
preemption of $o_3$ because it would be in
the interest of Peter to deviate at node $n_0$ towards $o_1$.

Then, knowing this, according to the second principle, Peter chooses $n_2$, since all
remaining outcomes in this subtree are better (3, 2, 5) than $o_1$ (1), which
is discarded.

At $n_2$, Mary has to take $n_4$ since $o_3$ was proven logically impossible (again,
her line of reasoning as a one-boxer rational player tells her that it would not be a
rational choice).

Then the application of the first principle here leads to the
preemption of $o_7$ because it would be in
the interest of Peter to deviate at node $n_4$ towards $o_5$.

Then, knowing this, according to the second principle, Peter chooses $n_6$, since all
remaining outcomes in this subtree are better (5) than $o_5$ (3), which
is discarded.

At $n_6$, Mary has to take $n_8$ since $o_7$ was proven logically impossible.

\begin{figure}[htbp]
\begin{center}
\includegraphics[width=8.4cm]{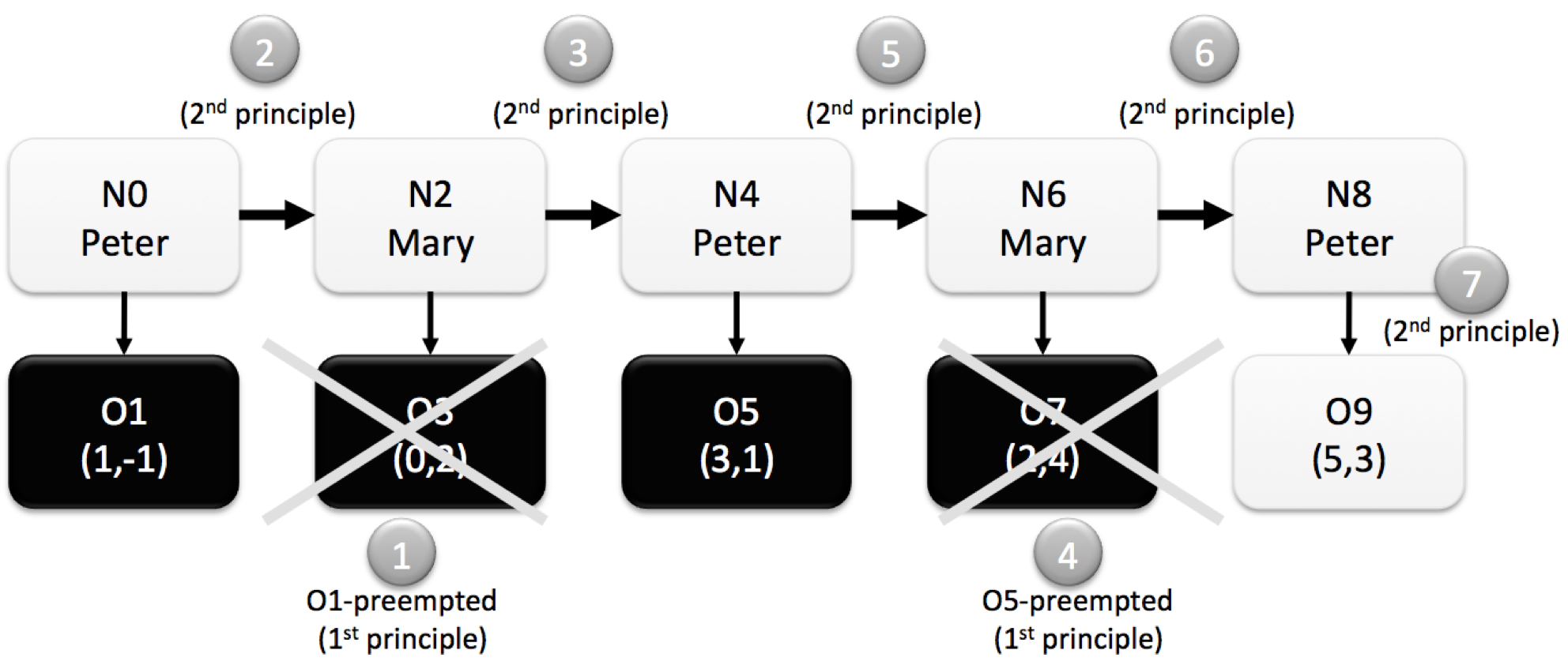}
\end{center}
\caption{The Perfect Prediction Equilibrium in the Take-Or-Leave game (n is
odd)}\label{Fig_TOL_Solved}
\end{figure}

At $n_8$, Peter chooses the remaining
outcome $o_9$, which is the Perfect Prediction Equilibrium (Fig.
\ref{Fig_TOL_Solved}).

\subsubsection{$\Gamma$-game (Fig. \ref{Fig_Gamma})}

In this game, the general technique for computing the Perfect Prediction
Equilibrium becomes visible.

We start at the root, where the first principle leads to the preemption of $o_4$
(Peter would deviate to $n_2$ where all payoffs are higher than -1) and then to
the preemption of $o_8$ (Peter would deviate to $n_1$ where all remaining
payoffs are now higher than 0).

\begin{figure}[htbp]
\begin{center}
\includegraphics[width=8.4cm]{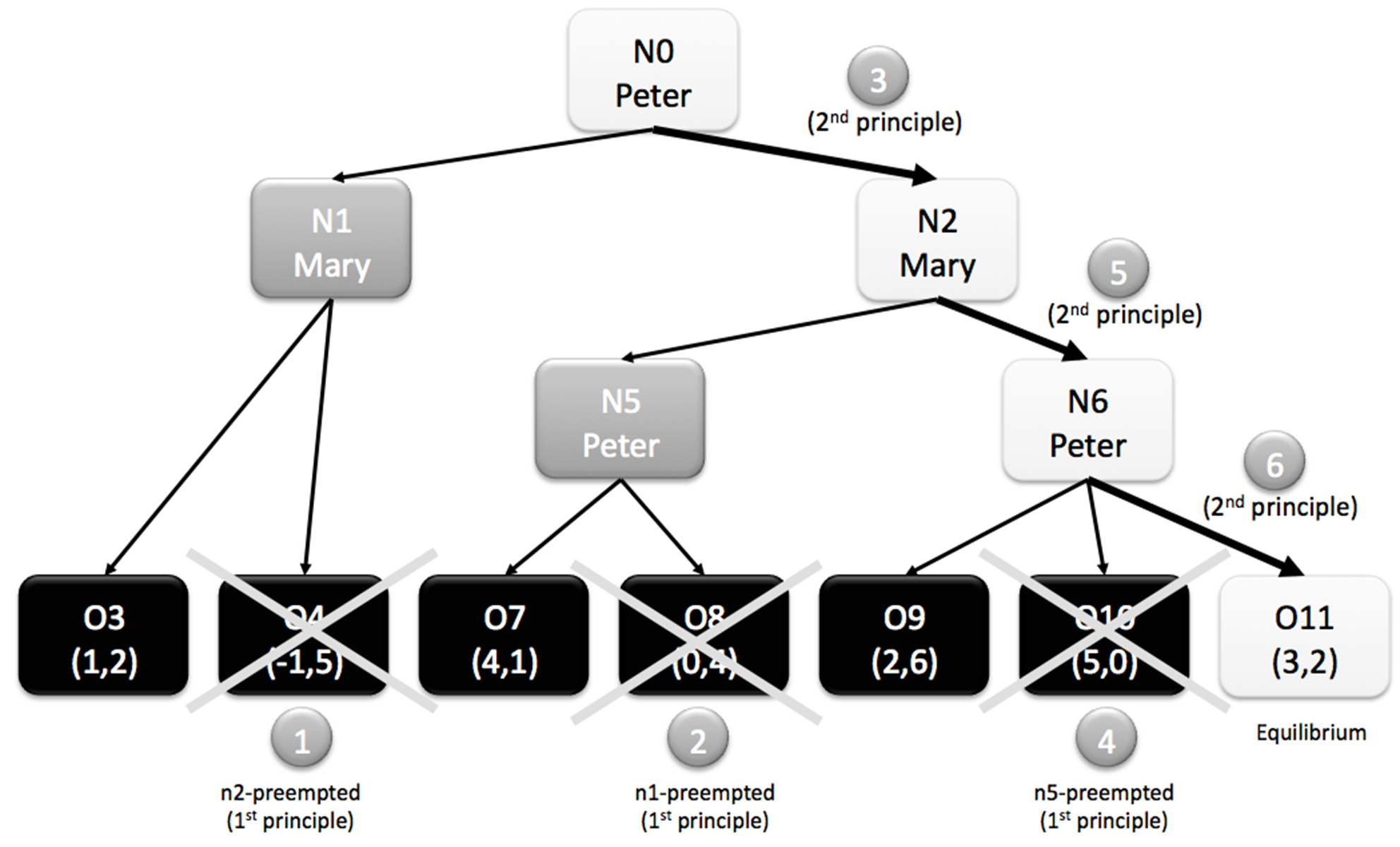}
\end{center}
\caption{The Perfect Prediction Equilibrium in the
$\Gamma$-game}\label{Fig_Gamma_Solved}
\end{figure}

Since the payoffs are no longer interlaced, we can now apply the second
principle. Peter chooses $n_2$ (remaining payoffs: 2, 3, 4, 5) over $n_1$ (1):
$n_2$ is now known to be reached by the game, $o_3$ is discarded (see Fig.
\ref{Fig_Gamma_Solved}).

At $n_2$ (where the only remaining outcomes are $o_7$, $o_9$, $o_{10}$ and
$o_{11}$), we apply the first principle with Mary: $o_{10}$ is preempted (Mary
would deviate to $n_5$). Then the second principle applies: Mary chooses $n_6$
(2, 3) over $n_5$ (1). $n_6$ is known to be reached by the game, $o_7$ is
discarded.

At $n_6$ (where the only remaining outcomes are $o_9$ and $o_{11}$), we apply
directly the second principle: Peter chooses $o_{11}$ (3) over $o_9$ (2).

The game reaches $o_{11}$. Conversely, the players, anticipating this outcome, will
play towards it: it is the equilibrium.

In fact, the algorithmic procedure will always be the same: identify the subgame
with the worst payoff and preempt its worst outcomes by making use of other
subgames as potential deviations.

\section{Construction of the Perfect Prediction Equilibrium}

\label{section-construction}

\subsection{A general definition}

In the former section, we introduced Perfect Prediction, the idea of preemption
and two principles, and constructed the PPE for some examples. We now define the
Perfect Prediction Equilibrium for all games in extensive form, without chance
moves and strict preferences.

We assume that both players are one-boxer rational. They consider that the solution
to the game is CK.

The application of the two principles detailed in Section \ref{section-perfect-prediction} allows the
elimination of all outcomes but one. The remaining, non-discarded outcome is the
Perfect Prediction Equilibrium. It is immune against the CK that
the players have of it. The two players, having anticipated it, have no interest
in modifying their choices and will play towards this outcome: it is a fixpoint.

Note that the definition of the Subgame Perfect Equilibrium in terms of
stability is very similar: it is an equilibrium which is reached when both
players have no interest in modifying their strategy, knowing the other player's
strategy, \emph{including what he would do at unreached nodes}. In our definition
however, we never consider unreached nodes: the Perfect Prediction Equilibrium
is stable with respect to the knowledge of itself as a path, and only of itself.
This implies that, as opposed to the Nash Equilibrium, there is no sense in
talking of ``equilibria in each subgame''. Actually the different subgames of
the tree disappear along the construction of the equilibrium: they are just used
as temporary possible deviations.
 
We believe that the Perfect Prediction Equilibrium fulfills a stability
condition which is simpler, more natural and more compact (self-contained) than
the Subgame Perfect Equilibrium. It may be regarded as both normative (because the Pareto optimality
is an incentive to adopt it
in the sense that it never reaches sub-optimal outcomes) and positive (it
describes game outcomes encountered in real life, like the assurance game).

\subsection{The construction of the Perfect Prediction Equilibrium}

We consider two-player\footnote{The formalism can very easily be extended
to any number of players. The main difference is that outcomes are labeled by
n-tuples of integers instead of couples.} finite games with perfect information,
without chance moves and with strict preferences.

\begin{definition}
(Labeled tree) A labeled tree is a tree whose root is $n_0$, and such that:

- each node $n$, labeled by a player $p_n$, has a set $F(n)$ of offsprings
(nodes or
outcomes).

- each leaf is an outcome $o$ labeled by a couple of integers ($a^{Peter}_o,
a^{Mary}_o$).

- all $(a^{Peter}_o)_o$ are distinct and all $(a^{Mary}_o)_o$ are distinct

\end{definition}

The label $p_n$ at node $n$ indicates whose turn it is to play. The offsprings
$F(n)$ of node $n$ are player $p_n$'s possible choices.

The label $(a^{Peter}_o, a^{Mary}_o)$ at outcome $o$ indicates how much Peter and
Mary get.

For convenience, nodes and outcomes are numbered ($n_0$, $n_1$ or $o_1$, $n_2$
or $o_2$, ...) so that it is possible to easily refer to them in the examples.

The last part of the definition corresponds to the assumption that each player
has a strict preference between any two outcomes.

We define $D(n)$ as the set of all descendants (nodes and outcomes) of node $n$; it is the
transitive and reflexive closure of the offspring-relation.

To compare outcomes, we write that $o_1 <_{p} o_2$ whenever $a^p_{o_1} <
a^p_{o_2}$. When the player is clear from the context, we will also say that an
outcome is better or worse than another one (meaning for the current player).

\begin{example}
In the Assurance game, we have two nodes, $n_0$ (the root) and $n_2$, and three
outcomes $o_1$, $o_3$ and $o_4$. The players are $p_{n_0}=$Peter and
$p_{n_2}$=Mary. F is defined as $F(n_0)=\{o_1, n_2\}$ and $F(n_2)=\{o_3, o_4\}$.
The payoffs are $(a^{Peter}_{o_1}, a^{Mary}_{o_1})=(0,0)$, $(a^{Peter}_{o_3},
a^{Mary}_{o_3})=(-1,2)$ and $(a^{Peter}_{o_4}, a^{Mary}_{o_4})=(1,1)$. Peter's
payoffs are all distinct, as well as Mary's payoffs. D is defined as
$D(n_0)=\{o_1, n_2, o_3, o_4\}$ and $D(n_2)=\{o_3, o_4\}$.
\end{example}

A very short summary of the construction of the Perfect Prediction Equilibrium
is the following: the reasoning is done by discarding all outcomes which cannot
be CK. We start at the root. The outcomes in all direct subtrees
but one are discarded, as well as some outcomes in the remaining subtree. The
next move has to be the root of the only subtree where at least one outcome
remains. Hence, the next move exists and is unique (Lemma \ref{lemplayersmove}).
We continue with the next move and so on, until an outcome is reached, which is
the outcome of the Perfect Prediction Equilibrium.

\subsubsection{Initialization, notations: $c_i$ and $I_i$}

An equilibrium is a path\footnote{We can either consider the equilibrium as the
outcome reached by the players, or as the path leading to it, which is
equivalent.} $(c_i)_{i=1..d}$ such that the first player starts at $c_1 = n_0$,
and at each step $i$, given $c_{i-1}$, the next move is $c_{i}\in F({c_{i-1}})$.

Formally :

\begin{itemize}
\item $c_1= n_0$ (the root)
\item $\forall i=1..d-1, c_i$ is a node
\item $\forall i=2..d, c_i \in F(c_{i-1})$
\item $c_d$ is an outcome
\end{itemize}

\begin{example} In the $\Gamma$-game, the equilibrium is $(c_1, c_2, c_3, c_4)$
where $c_1=n_0, c_2=n_2, c_3=n_6, c_4=o_{11}$.
\end{example}
At each step $i$ we are also given $I_{i-1}$, a subset of the set $D(c_{i-1})$
of the outcomes descending from the previous move $c_{i-1}$. This set $I_{i-1}$
is the set of all outcomes that are immune against their knowledge up to
$c_{i-1}$ (technically speaking, they have not been discarded yet: the second
principle has only been applied up to $c_{i-1}$, their common ancestor, and they
could not be preempted with the first principle at any node earlier on the
path).

\begin{example} In the $\Gamma$-game, the sequence would be $I_1=\{\textbf{all
outcomes}\}$, $I_2=\{o_7, o_9, o_{10}, o_{11}\}$, $I_3=\{o_9, o_{11}\},
I_4=\{o_{11}\}$. $o_7$ is in $I_2$ because it is not discarded, meaning that if
Peter at $c_1=n_0$ anticipates $o_7$, he will not deviate. $o_8$ does not make
it in $I_2$ because it is preempted by Peter at $n_0$ according to the first
principle. $o_3$ is not in $I_2$ either since it is discarded by the second
principle, as well as all outcomes not in the $n_2$ subtree.
\end{example}

A node that is not in $I_{i-1}$ cannot be commonly known to be reached by the
game, either because it was discarded by the second principle, or because it
would bring about a deviation in the past according to the first principle. Note
that sometimes, when the outcomes in two subtrees are not interlaced, it can be
ambiguous which principle to apply to discard some outcomes: when the second
principle potentially selects a subtree and discards the other outcomes, then
the first principle could also be used instead to preempt the very same outcomes
with this subtree. This leaves room for a philosophical interpretation
(preemption or rational choice), but does not change what outcomes are
discarded. This is why, in this section, we will apply the first principle
whenever it applies, and then the second principle, but we will push forward the
word ``discard'' instead of ``preempt'' to leave open which principle is used in
the interpretation.

\begin{example} In the $\Gamma$-game, one could also argue that $o_3$ is not in
$I_2$ by invoking the first principle: $o_3$ is preempted by the $n_2$ move. Even
with this interpretation, it changes nothing to the fact that all outcomes not
in the $n_2$ subtree are discarded.
\end{example}

In other words, the outcomes that are not in the set $I_{i-1}$ are not to be
taken into account anymore: they have been proven not to fit the definition of
the equilibrium. The other partially fit the definition: they are immune against
their knowledge at least up to $c_{i-1}$, which means that players will play
towards this outcome at least until $c_{i-1}$ is reached, but could still
deviate later.

At the beginning $I_1=\{$all outcomes$\}$, which means that no outcome has been
discarded yet: $c_1$ is the root and is always reached because the game starts
here (the causal bridge is secured).

At each step, we discard outcomes (with the first principle, or possibly the
second principle for the very last discarding, which is technically equivalent)
until all subtrees but one are completely discarded, and finish with the second
principle which allows us to move on to the next node and the next step.

\subsubsection{Newcombian States}

At step $i$, the last move $c_{i-1}$ and the remaining outcomes $I_{i-1}$ are
given. For example at step $2$, the last move is the root ($c_1 = n_0$) and the
remaining outcomes are all the outcomes ($I_1 = \{$all outcomes$\}$), and we are
looking for $c_2$ and $I_2$.

We will now define Newcombian States\footnote{Newcombian States can be seen as a
the mathematical counterpart of a player's Perfect Prediction skills. This name
has been chosen as a tribute to Newcomb's Problem (see Section \ref{section-newcomb}).
To reach the Perfect Prediction Equilibrium, players reason as
one-boxers.} for the player $p_{c_{i-1}}$ (the current player) so as to
recursively determine the outcomes that are not immune against their common
knowledge because of a deviation at the node $c_{i-1}$. These outcomes are
discarded at this step.

A Newcombian State has a pure part (corresponding to a possible move) and a
discarding part (a Newcombian State which invalidates some outcomes as
equilibrium candidates, with one of the two principles). It can be thought of as
a bundle containing a move (its ``pure part'') together with the knowledge and
the proof (its ``discarding part'') that some of its descendant outcomes are
impossible. The proof that such an outcome is impossible consists of a witness:
the move in the discarding part
\begin{itemize}
\item to which the player $p_{i-1}$ would deviate according to the first
principle if the outcome were predicted.
\item or to which the player $p_{i-1}$ actually goes according to the second
principle.
\end{itemize}

\begin{definition}
(Newcombian State, pure part, discarding part) At step $i$, a Newcombian State
of order $k$, with $k\geq1$,
is an element $\eta=(\eta_1,..,\eta_k) \in {(F(c_{i-1}))}^k$, constituted by
offsprings of the last move, such
that any two consecutive elements are different:

\begin{equation}
\eta=(\eta_1,..,\eta_k) \in {(F(c_{i-1}))}^k:  \eta_j \neq \eta_{j+1},
\quad j=1..k-1
\end{equation}

The pure part of a Newcombian State is its first element:
$\mathbf{pure}(\eta)=\eta_1 \in F(c_{i-1})$.

The discarding part of a Newcombian State of order $k>1$ is
the element of $({F(c_{i-1})})^{k-1}$ composed by the $k-1$ last
elements of $\eta$: 
$$\mathbf{discard}(\eta)=(\eta_2,..,\eta_k) \in {(F(c_{i-1}))}^{k-1}$$.

\end{definition}

\begin{example}
In the $\Gamma$-game, $(n_1,n_2)$ is a Newcombian state of order 2. Its pure
part is $n_1$ and its discarding part is $n_2$. $n_2$ is a witness that a
descendant ($o_4$) of $n_1$ is impossible, because it is discarded by the first
principle ($o_4$ is preempted by $n_2$).

Likewise, $(n_2, n_1, n_2)$ is a Newcombian state of order 3. Its pure part is
$n_2$ and its discarding part is $(n_1, n_2)$. The latter is a witness that a
descendant ($o_8$) of $n_2$ is discarded by the first principle (it is preempted
by $n_1$ knowing that $o_4$ is impossible).

Finally, $(n_1, n_2, n_1, n_2)$ is a Newcombian state of order 4. Its pure part
is $n_1$ and its discarding part is $(n_2, n_1, n_2)$. The latter is a witness
that a descendant of $n_1$ ($o_3$) is impossible. In this case, it is open
whether the second principle (Peter chooses $n_2$, knowing that $o_8$ and $o_4$
are impossible, which discards all outcomes in other subtrees) or the first
principle ($o_3$ is preempted by $n_2$ knowing that $o_8$ and $o_4$ are
impossible) is applied, liked we mentioned above. Whichever principle is used,
$o_3$ is discarded.
\end{example}

A Newcombian State of order 1 is a pure Newcombian State: no outcome has been
discarded yet (unless it was at a former step, but we said that outcomes
invalidated at a former step are out of the race anyway).

\begin{example}
$(n_2)$ is a pure Newcombian state, i.e., in which no outcomes have been
discarded yet.
\end{example}

\subsubsection{Target function and discarding}

Each Newcombian State is associated with the subset of the descendants of its
pure part that are not discarded by its discarding part. This is defined
recursively by making appeal to the target function $T_i$ in the following way:

\begin{definition}
\label{defpreemption}
(Target function and Discarding) At step $i$,

- if $\eta$ is a first-order Newcombian State, i.e. $\eta=(\eta_1)$,
then $T_i(\eta)$ is the set of outcomes that are among the descendants
of $\eta_1$ and that were not discarded before step $i$: 
$$T_i(\eta)=I_{i-1} \cap D(\eta_1)$$

- if $\eta$ is a Newcombian State of order $k>1$, i.e.
$\eta=(\eta_1,p)$, then $T_i(\eta)$ is the set of outcomes that are
among the descendants of $\eta_1$ and that were not discarded before step $i$,
\textit{excluding} outcomes worse than the
worst\footnote{An outcome is said \emph{worse} than
another outcome if and only if it gives a lower payoff with respect
to the current player $p_{c_{i-1}}$.} outcome in $T_i(p)$ if $T_i(p)$ is not
empty.

Every outcome that is excluded that way is
called $p$-\textbf{discarded}. Any outcome in $T_i(\eta)$ is said
$\mathbf{\eta}$\textbf{-targeted}. If we denote the set of the $p$-discarded
outcomes with
$$P_i(p) = \{o\in I_{i-1} \setminus D(\textbf{pure}(p)) | T_i(p) \ne \emptyset
\wedge \forall o' \in T_i(p),o <_{p_{c_{i-1}}} o' \}$$

then
$$T_i(\eta) = (I_{i-1} \cap D(\eta_1)) \setminus P_i(p)$$

which defines a recursion on $k$.

\end{definition}

Note the terminology: an \emph{outcome} is discarded by a \emph{Newcombian
State}.

The operation of discarding corresponds to the player's behavior described in
the principles:
\begin{itemize}
\item With the first principle, if, predicting outcome $o$ will happen, she
moves to a subtree with outcomes that are all better than $o$, then $o$
is discarded and is no longer an equilibrium candidate.

\item With the second principle, the player selects a move, which discards any
outcome in the other subtrees.
\end{itemize}

\begin{example}
For example in the $\Gamma$-game, at the root, i.e. step 2, the target functions
for the pure Newcombian states are $T_2(n_1)=\{o_3, o_4\}$ and $T_2(n_2)=\{o_7,
o_8, o_9, o_{10}, o_{11}\}$ (they contain all descendants). The corresponding
discarding functions are $P_2(n_1)=\emptyset$ and $P_2(n_2)=\{o_4\}$. Knowing
this, it is possible to compute the target functions of the Newcombian states of
order 1: $T_2(n_2, n_1)=T_2(n_2) \setminus P_2(n_1)=T_2(n_2)$ and $T_2(n_1,
n_2)=T_2(n_1)\setminus P_2(n_2)=\{o_3, o_4\} \setminus\{o_4\}=\{o_3\}$.
\end{example}

There are two extreme cases for discarding:
\begin{itemize}
\item If $T_i(p)=\emptyset$, or if there is no $\eta_1$-targeted outcome worse
than the worst
$p$-targeted outcome, then $T_i(\eta)=T_i(\eta_1)$. Discarding
$\eta_1$-targeted outcomes by $p$ is ineffective.

\item If all $\eta$-targeted outcomes are worse than the worst
$p$-targeted outcome, then $T_i(\eta)=\emptyset$. This time,
discarding the descendance of $\eta_1$ by $p$ is at its highest
efficiency, since $p$ discards every $\eta_1$-targeted outcome.
$\eta$ is called a degenerate state. A degenerate state cannot discard any
outcome.

It is only for a degenerate state $\eta$ that it is ambiguous which principle is
being applied: the first one (the outcomes in $P_i(p)$ are preempted) or the
second one (the player moves rationally to the pure part of $\eta$). For
non-degenerate states, only the first principle makes sense.

\end{itemize}

\subsubsection{Example}

\begin{figure}[ht]
\centering
\includegraphics[width=8.4cm]{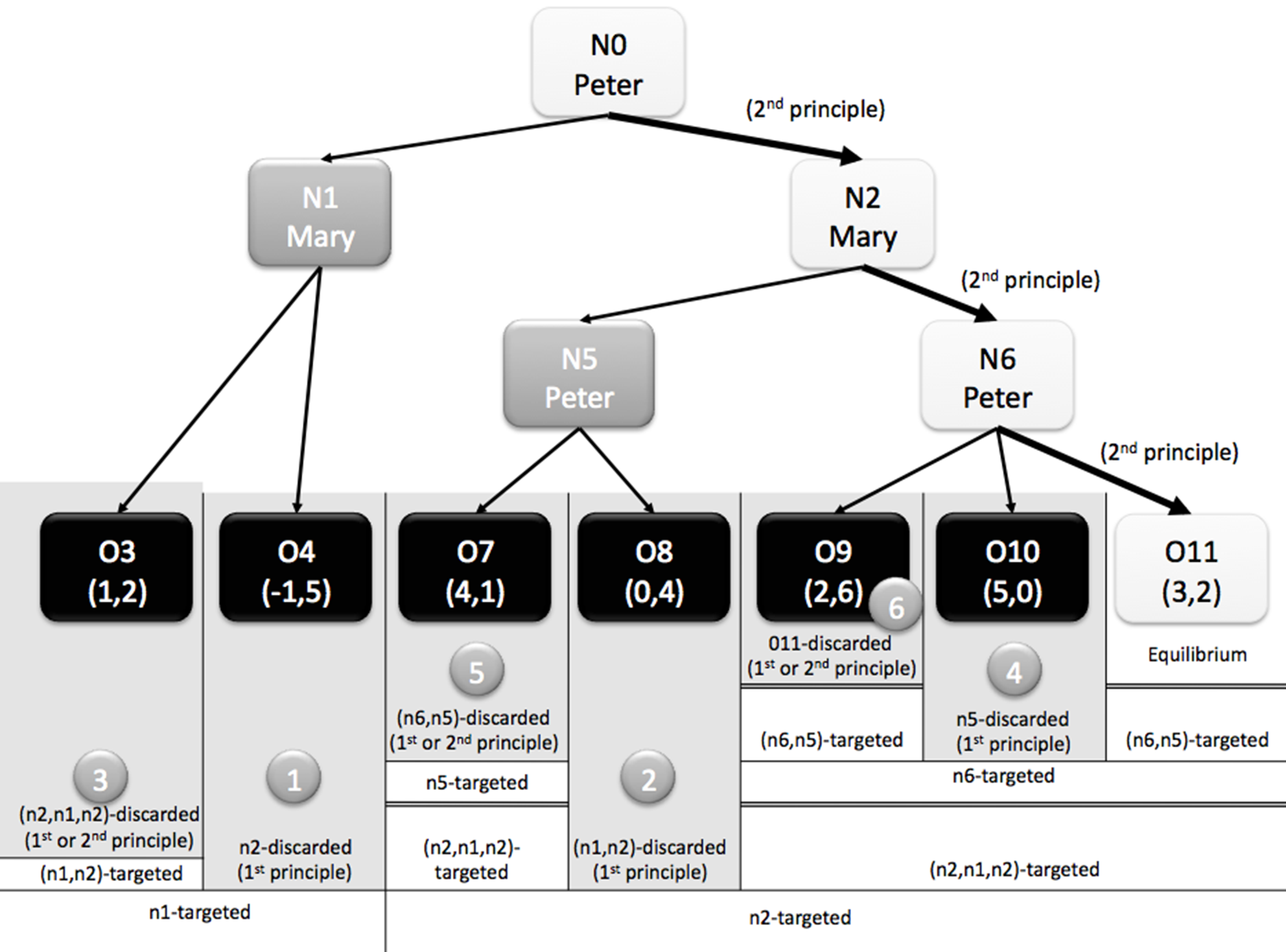}
\caption{The $\Gamma$-game with Newcombian states and their targeted/discarded outcomes at
the root.}
\label{Fig_Gamma_Target}
\end{figure}

At step $1$ of game $\Gamma$, we have $c_1=n_0$ and $I_1=\{$all outcomes$\}$
(initial values).

At step $2$, we compute the target functions of the Newcombian States:
\begin{eqnarray*}
 T_2(n_1)&=&\{o_3, o_4\} \\
T_2(n_2)&=&\{o_7, o_8, o_9, o_{10}, o_{11}\}\\
&&\\
P_2(n_1)&=&\emptyset\\
T_2(n_2, n_1)&=&T_2(n_2)\\
P_2(n_2)&=&\{o_4\}\\
T_2(n_1, n_2)&=&T_2(n_1)\setminus P_2(n_2)=\{o_3, o_4\}
\setminus\{o_4\}=\{o_3\}
\end{eqnarray*}

\begin{eqnarray*}
P_2(n_1,n_2) & = & \{o_8\} \\
T_2(n_2, n_1, n_2) & = & T_2(n_2) \setminus P_2(n_2)\\
& =&\{o_7, o_8, o_9, o_{10}, o_{11}\} \setminus\{o_8\}=\{o_7, o_9, o_{10}, o_{11}\}\\
P_2(n_2,n_1,n_2)&=&\{o_3, o_4\}\\
T_2(n_1, n_2, n_1,n_2)&=&T_2(n_1) \setminus P_2(n_2,n_1,n_2) = \{o_3, o_4\}
\setminus\{o_3, o_4\} = \emptyset.
\end{eqnarray*}

The entire reasoning (also at $n_2$ and $n_6$) is indicated on Fig.
\ref{Fig_Gamma_Target}, where the order of the sub-steps is indicated by the numbers
in gray circles. Sub-steps 1, 2 and 3 are performed by Peter at $n_0$. One
notices that, by discarding all outcomes we can, we are left with outcomes
$o_7$, $o_9$, $o_{10}$ and $o_{11}$, and all of them are descendants of $n_2$,
which is chosen by Peter ($c_2 = n_2$ and $I_2=\{o_7, o_9, o_{10},o_{11}\}$).
Sub-steps 4 and 5 are then performed by Mary at $n_2$ and Sub-step $6$ at $n_6$
by Peter. Actually, it always works in the same way, at each step and for any
game: we can discard the outcomes of all subtrees but one, as demonstrated in
the next part.

\subsubsection{The current player's move}

In the example, we saw that after discarding as many outcomes as possible at the
root, only outcomes in exactly one subtree remained - this means that this
subtree corresponds to the player's next move. We will now show that at any node
reached by the game, the current player's move exists and is unique.

Having constructed all Newcombian States at step $i$, a certain amount of
outcomes, having been discarded, are impossible. We consider all the outcomes
that are still available after this: these are the outcomes that are immune
against their knowledge up to the current step.

The set $I_i \subset I_{i-1}$ of all remaining outcomes is given by:
$$I_i=I_{i-1} \setminus (\bigcup_{\eta} P_i(\eta))$$

Note that this infinite union is merely a notation, and represents in fact a
finite union, since all the $P_i(\eta)$ are subsets of $I_{i-1}$ (and the set of
subsets of a finite set is finite).
Nevertheless, despite being practical, this notation does not allow to create
any algorithm for computing the equilibrium.
We will present another characterization of the current player's move
which permits this in Section \ref{section-explicit-player-move}.

\begin{example}
In the $\Gamma$-game, after reasoning at the root, the set of remaining outcomes
is $I_2 = I_1 \setminus (P_2(n_1) \cup P_2(n_2) \cup P_2(n_1, n_2) \cup P_2(n_2,
n_1) \cup ...) = \{o_7, o_9, o_{10}, o_{11}\}$. Note that it is impossible to
algorithmically compute the infinite union with this formula, but the result is
obtained by noticing that all other outcomes have been discarded, and these four
remaining outcomes cannot be discarded by any of the two principles: they are
all better for Peter than any outcome in any other subtree.
\end{example}

\begin{lemma}
\label{lemplayersmove} Existence and uniqueness of current player's move

1. (Existence of the current player's move) There is at least one outcome which
has not been discarded during the process:
$$I_i =I_{i-1} \setminus (\bigcup_{\eta} P_i(\eta)) \ne \emptyset $$
2. (Uniqueness of the current player's move) All remaining outcomes are the
descendants of one unique offspring of the last player's move $c_{i-1}$:
$$\exists ! f \in F(c_{i-1}), I_i \cap D(f) \ne \emptyset$$
and this very offspring is the current player's move $c_i$:
$$\{c_i\}\equiv\{f \in F(c_{i-1}), I_i \cap D(f) \ne \emptyset\}$$

\end{lemma}

\begin{example}
If we continue with the example above, we see that all remaining outcomes ($o_7$,
$o_9$, $o_{10}$, $o_{11}$) are descendants of $n_2$. This implies that $c_2=n_2$.
\end{example}

Now that all outcomes in all subtrees but one have been discarded, the game
reaches $c_i$: whatever outcome in $I_i$ is anticipated, the players play
towards $c_i$. Hence, the reasoning can continue at this node, at which further
outcomes can be discarded.

\subsubsection{End of recursion - Definition and Theorem}

When the path reaches an outcome $c_d$ ($I_d=\{c_d\}$), the induction stops,
leaving one non-discarded outcome.

\begin{example}
In the $\Gamma$-game, at step 3, the game reaches $c_3=n_6$ with the remaining
outcomes $I_3=\{o_9, o_{11}\}$. At step 4, it reaches $c_4=o_{11}$ with
$I_4=\{o_{11}\}$. The induction stops here: $o_{11}$ is reached by the game.
($n_0, n_2, n_6, o_{11}$) is the equilibrium: under the assumption of total
transparency, the players will reach it; even though they know it in advance,
they never deviate. Actually, the fact that they know it in advance \emph{causes} them to reach it.
\end{example}

\begin{definition}(Perfect Prediction Equilibrium)
\label{defppe}
In a game in extensive form represented by a finite tree, with perfect
information, without chance moves and with strict preferences between the
outcomes, under the assumption of total transparency, a Perfect Prediction
Equilibrium is a path leading to a non-discarded outcome, i.e., an outcome which
is immune against the CK that the players have of it: the players,
anticipating it, have no interest in modifying their choices and do play towards
this outcome. It is a fixpoint, in that it follows causally from its prediction.
\end{definition}

\begin{theorem}(Existence and uniqueness of the Perfect Prediction Equilibrium)
\label{thmppe}
In a game in extensive form represented by a finite tree, with perfect
information, without chance moves and with strict preferences between the
outcomes, there is one unique Perfect Prediction Equilibrium.

\end{theorem}

\subsection{Pareto optimality}

The Perfect Prediction Equilibrium has the additional property of always being
Pareto-optimal. This concept is of crucial importance in economics. In our case,
it means that when the PPE is reached, the payoffs are distributed so that no
other equilibrium can make a player better off without making the other player
worse off.

\begin{theorem}
\label{thmpareto}
The outcome of the Perfect Prediction Equilibrium is Pareto-optimal among the
outcomes of the game.
\end{theorem}

Note that, although the PPE is always Pareto-optimal, it is not always a Pareto
improvement of the SPE (for a counterexample, see the complete analysis of biped
games in the Appendix, case ``$1\neq$'' with $d=1$ and $f=0$).

\section{Related Work}
\label{section-related-work}

\subsection{The Backward Induction Paradox}

\label{section-backward-induction-paradox}

In the last decades, there has been a debate around the so-called Backward Induction Paradox: CK of
rationality is also assumed at nodes that are known to be not
reached given this very CK of rationality. This points out that in some way, the SPE
is not completely transparent to itself: the players do know their respective strategies, and though
cannot be fully -- that is, at all nodes in the tree -- cognizant of the final equilibrium determined by these two
strategies.

One of the motivations behind the Perfect Prediction Equilibrium is to avoid the Backward Induction Paradox (BIP). In this section, we show how literature relates the BIP to the assumption of counterfactual independence and hence,  why giving up this assumption circumvents the BIP.

Two kinds of BIPs can be found in literature.

\citet{BIP} describe an empirical BIP with the example of the
prisoner's dilemma (a game with imperfect information): ``There is a well-known argument --
the backward induction argument -- to the effect that, in such a sequence, agents who are
rational and who share the belief that they are rational will defect in every round. This argument holds however large
$n$ may be. And yet, if $n$ is a large number, it appears that I might do better to follow a strategy such as
tit-for-tat. [...] This is the backward induction paradox.''

\citet{Baltag2009} describe a more fundamental, logical BIP as follows: ``in order even to start the reasoning, a player assumes that (common knowledge of, or some form of common belief in) \emph{rationality}' holds at all the last decision nodes (and so the obviously irrational leaves are eliminated); but then, in the next reasoning step (going backward along the tree), some of these (last) decision nodes are eliminated, as being incompatible with (common belief in) \emph{rationality}!'' 
\citet{PRREFG} showed that in the Take-or-Leave game, CK of rationality is self-contradictory at each node outside of the equilibrium path.

In the last two decades, the Backward Induction Paradox has been the object of an uninterrupted debate. On the one side, \citet{RABICKR} proved that ``common knowledge of rationality (CKR) implies [that players are reasoning by] backward induction.'' (although he does not ``claim that CKR obtains or should obtain''). On the other side, Stalnaker states that ``Common knowledge of substantive\footnote{At all vertices in the game. This is the definition used by \citet{Halpern2001}.} rationality does not imply the backwards induction solution'' \citep{Halpern2001} and argued \citep{Stalnaker1998} that Aumann's argument ``conflates epistemic\footnote{We call this counterfactual independence.} and causal independence, implicitly making a strong epistemic independence assumption which it explicitly rejects.'' The opposition to Aumann's claim was supported by \citep{KBNBI} and
\citep{KBRBI}.

\citet{Halpern2001} showed that both sides are right, in that ``the key difference between Aumann and Stalnaker lies in how they interpret'' the counterfactual' statement ``for all vertices $v$, if the player were to reach vertex $v$, then the player would be rational at vertex $v$'' (substantive rationality), i.e., whether it is a material\footnote{i.e., equivalent to ``the players do not reach vertex $v$, or they are rational at vertex $v$''} or a counterfactual implication\footnote{i.e., ``if [the player] were to actually reach $v$, then what he would do in that case would be rational'' \citep{Halpern2001}}.

\citet{Baltag2009} go in another direction by allowing players to have``a moment of temporary irrationality.''

We believe that the Backward Induction Paradox is due to the lack of transparency arising from the widespread assumption that present decisions are counterfactually independent of past moves, an assumption also made explicitly by \citet{Baltag2009}: ``players have no non-trivial ``hard'' information about the outcomes [...]: they cannot foretell the future, cannot irrevocably know the players' freely chosen future moves (though they do irrevocably know the past, and they may irrevocably know the present [...])''.

Perfection prediction theory aims at giving an alternate game-theoretical solution concept with the exact opposite assumption: there is CK of the outcome of the game, and the past is counterfactually dependent on this outcome. There is also CK of rationality on the equilibrium path, and no need for it outside of this path (no counterfactuals on (un)reached nodes, no substantive rationality).

This happens to lead to cooperative, Pareto-optimal behavior. Hence, the Backward Induction Paradox, in the two formulations given above, does not manifest itself in the PPE theory.

Supporting the possibility of having an alternate theory of rational choice,
\citet{DKCDDB} argues that ``when one says ``rational choice theory'', it sounds
as if only one theory or model of choice could qualify. How could two distinct
theories or models both be rational? But people behave in different ways,
depending on the specific context and the more general social situation, and I
see no reason to privilege one universal model of behavior with the adjective
rational. To do so is, if not demonstrably silly, at least demonstrably
misleading.''

\subsection{Super-Rationality}

\citet{Hofstadter1983} introduced the concept of super-rationality like so: "Super-rational thinkers, by recursive definition, include in their calculations the fact that they are in a group of super-rational thinkers." Super-rationality is defined for symmetric games. We believe that the one-boxer rationality underlying the PPE is the equivalent of super-rationality for games in extensive form, as it shares the core idea of transcending rationality, in that players include the knowledge thereof in their calculations.

\citet{Shiffrin2009} argue that rationality is "not a normative concept but rather a social
consensus of a sufficiently large proportion of humans judged to be sufficiently expert." They point out
that normative game theory reaches non-(Parent)-efficient optima and initiated research to suggest
an alternate equilibrium concept that aims at being Pareto-optimal and that differs from the Nash equilibrium. They explicitly point to Hofstadter's super-rationality as a starting point. Also, they compare their approach to a single player's playing against himself under the effect of the Midazolam drug, which makes him forget his past decisions while preserving his reasoning abilities.
It can be taken from the draft that, unlike the PPE, no preemption structure is considered. Also, they are facing an exponential explosion of the search space for more than 2 players. The PPE scales to any number of players with the same algorithms.

\subsection{Forward Induction}

\label{section-forward-induction}

The Perfect Prediction Equilibrium is \emph{de facto} built with a forward
induction, since its construction begins at the root and ends at an outcome. Note that, in game theory, the
expression ``Forward Induction'' carries much more meaning than the fact that
the reasoning follows a path from the root to an outcome.

Although in the past many Forward Induction Equilibria have been designed,
according to \citet{GW}, ``the literature provides no formal definition of
Forward Induction.''

\bigskip
We now compare our equilibrium with others found in
literature by first looking at differences and then at similarities.

First of all, we limit ourselves to games with \emph{perfect information},
meaning that information sets are all singletons, so that beliefs (as they are
defined in the Sequential Equilibria of \citet{KW}) are the trivial function 1
on a singleton.

Additionally, Forward Induction is often used with mixed strategies leading to
probability distribution over the game outcomes. Usually, Forward Induction
seems to select an equilibrium among the Sequential Equilibria as defined by
\citet{KW}. In our work we only consider pure strategies, with pure game
outcomes.

Finally, there is the issue of what happens outside of the path. It has already
been addressed, for example in the Weakly Sequential Equilibrium by \citet{KW},
where strategies are required to be optimal only on non-excluded information
sets (but are defined also on information sets excluded by these very
strategies).
\citet{CK} introduced signaling games, in which they also address this issue:
``If one can restrict the out-of-equilibrium beliefs (or hypotheses) of B, one
can sometimes eliminate many of the equilibria.'' We find the same idea in the
work of \citet{BS}: ``This section presents an equilibrium concept that refines
the set of sequential equilibria in signaling games by placing restrictions on
off-the-equilibrium-path beliefs''.
In our case, pure strategies are simply \emph{not defined at all} outside the
equilibrium path. Actually, outside the equilibrium path, it is not even assumed
that players are rational.

\bigskip
In spite of these differences (perfect information, pure strategies, no
definition outside of the equilibrium path), we can find some interesting
similarities with existing Forward Induction Equilibria.

It is commonly assumed that past play was rational, as opposed to backward
induction, where future play only is supposed to be rational. See for example
\citet{GW}: ``Kohlberg and Mertens label this result Forward Induction but they
and other authors do not define the criterion explicitly. The main idea is the
one expressed by Hillas and Kohlberg in their recent survey: ``Forward induction
involves an assumption that players assume, even if they see something
unexpected, that the other players chose rationally in the past'', to which one
can add that `and other players will choose rationally in the future.' This is
implicit since rationality presumes that prior actions are part of an optimal
strategy.'' We also assume rationality of other players in the past, but even
more than that, we assume that the past has a power of preemption. Technically,
in occurring time for example, it could mean that if a player (she) chooses a
move towards a preempted outcome, knowing that the other player (he) anticipated
this outcome, this contradicts his rationality in the past, hence, she cannot
choose such a move.

Preemption has also already been indirectly used in Forward Induction: e.g., by
\citet{CK}: ``Again envisaging the thought process of the players, we have in
mind something like Kohlberg and Merten's process of forward induction: will
some type of player A, having arrived introspectively at restrictions in B's
beliefs (hence B's conceivable actions), see that deviation will lead to a
higher payoff than will following the equilibrium?''

CK of the equilibrium has already been considered as well:
\citet{CK}: ``An equilibrium is meant to be a candidate for a mode of
self-enforcing behavior that is common knowledge among the players. (Most
justifications for Nash equilibria come down to something like this. See, for
example Aumann [1987] or Kreps [forthcoming]). In testing a particular
equilibrium (or equilibrium outcome), one holds to the hypothesis that the
equilibrium (outcome) is common knowledge among the players, and one looks for
``contradictions''.'' This is exactly what we do in PPE: it is the only
equilibrium stable against its CK. In our case, it is unique and
all other paths lead to contradictions.

\bigskip
Note that \citet{GW} also give a proposal for a Forward Induction criterion with
mixed strategies and without the perfect information assumption: ``Definition
3.4 (Forward Induction). An outcome satisfies forward induction if it results
from a weakly sequential equilibrium in which at every relevant information set
the support of the belief of the player acting there is confined to profiles of
nature's strategies and other players' relevant strategies.'' Although the
framework is different, it can be interesting to ask whether the PPE fulfills
this criterion, knowing that we are in the special case where
we have perfection information and pure strategies which are
a special case of mixed strategies, and where the 'knowledge' of each player can be
seen as beliefs with probability 1.

The Sequential Equilibrium is close to the Subgame Perfect Equilibrium: in
\citep{KW}, Proposition 3 says: ``if $(\mu, \pi)$ is a Sequential Equilibrium,
then $\pi$ is a Subgame Perfect Nash Equilibrium.'' Since the PPE is not always
an SPE, it is not always a Sequential Equilibrium either.

Now, the Weakly Sequential Equilibrium is more general than the Sequential
Equilibrium, in that players' strategies do not need to be optimal at
information sets excluded by the equilibrium strategy. Unfortunately, this is
not enough: a PPE is not always a WSE either. For example, in the assurance
game, the only WSE outcome is (0,0) whereas the PPE leads to its Pareto
improvement (1,1). What happens with the WSE is that (1,1), which is the PPE
outcome, is already discarded at the beginning of the reasoning by the second
player who prefers (-1,2), and the first player then chooses (0,0) by backward
induction: there is no notion of preempting (-1,2), which would influence the
choice of the second player.

\bigskip
Hence, the Perfect Prediction Equilibrium is not a refinement of any of the
above equilibria. While using preemption and CK of the equilibrium
outcome, which are not new and have already been addressed in Forward Induction
literature, it deviates from the ideas behind the Subgame Perfect Equilibrium.

Although we can find the same intuitions in the Forward Induction
literature and in the construction of the PPE, the literature still (implicitly)
assumes counterfactual independence of a player's move from the past,
whereas we are using the concept of Perfect Prediction.

Starting with the premise that the game is played by agents for whom the world
is totally transparent, we established that
exactly one equilibrium, the Perfect Prediction Equilibrium, fulfills this
requirement.

\section{Conclusion}

We introduced an alternate solution concept, the Perfect Prediction Equilibrium, which is reached
by players that have a different form of rationality than that used in Subgame Perfect Equilibrium.
It accounts for the behavior of players that would pick one box in Newcomb's problem, as opposed to rational players in the Nash sense, that would pick two boxes with a dominant strategy argument.

One-boxer-rational players seem to get a reward over two-boxer-rational players, as the former can only reach Pareto-optimal outcomes.

Table \ref{spe-ppe-summary} summarizes the differences between the SPE paradigm and the PPE paradigm.

\begin{table}
\caption{A summary of the differences between the SPE and PPE paradigms}
\label{spe-ppe-summary}
\begin{tabular}{|c|c|c|}
\hline
& Subgame Perfect & Perfect Prediction  \\
& Equilibrium & Equilibrium \\
\hline
Form of the Game & Extensive & Extensive \\
\hline
Perfect Information & Yes & Yes \\
\hline
Number of players & Any & Any \\
\hline
Newcomb Choice & Two boxes & One box\\
\hline
Prediction Model & Ad-Hoc Prediction & Perfect Prediction \\
 accounting for & (``could have & (``would also \\
counterfactuals & been wrong'') & have been right'') \\
\hline
Relationship with the Past & Cournot-like & Stackelberg-like \\
\hline
Reasoning & Backward Induction & Forward Induction\\
\hline
Indifference between payoffs & Allowed & Not allowed in principle \\
\hline
Existence & Always & Always \\
\hline
Uniqueness & Always & Always \\
\hline
Optimality & -  & Pareto \\
\hline
Corresponding  & Nash & Superrationality (not established/\\
Normal Form &  (subsuming) & conceptually similar) \\
\hline
\end{tabular}
\end{table}

As a conclusive remark, it is important to note that, for the equilibrium to be reached, it does not require
that Perfect Prediction holds. It suffices that the one-boxer rational players believe that it does. Furthermore, the algorithms are simple and efficiently computable. The PPE can also be used by pragmatic two-boxer-rational players to mutually agree on a settlement contract to get a more efficient outcome than they would otherwise. Hence, even if an agent who feels very strong about
two-boxer-rationality will not find one-boxer-rationality reasonable, they cannot deny that its benefits are quite reasonable.

\section{Acknowledgments}

We are thankful to Alexei Grinbaum and Bernard Walliser for their precious
advice.

\bibliographystyle{spbasic}

\bigskip

\section{Annex - Preemption Structure and Algorithms}

\label{section-preemption-structure}

We proved that the current player's move exists and is
unique, but we did not give an explicit way of determining how to compute it. In
order to introduce algorithms, we need a further analysis of the preemption
structure. After giving an explicit construction of the equilibrium, we derive
two algorithms.

\subsection{Preemption structure analysis - the explicit construction of the
PPE}

In this part, we first introduce Newcombian Classes for a given step $i$ and
then give an explicit characterization of the player's move at the corresponding
node.

\subsubsection{Newcombian Classes}

Two Newcombian states with the same targeting function can be considered
equivalent.

\begin{definition}
(target-equivalent, Newcombian Class) With respect to step $i$, two Newcombian
States $\eta$ and $\eta'$ are
target-equivalent if they target the same
outcomes: $T_i(\eta)=T_i(\eta')$.

This defines an equivalence relation on the set of all Newcombian States
relative to the same move:
it is reflexive, symmetric, and transitive.
The equivalence class of a Newcombian State $\eta$ is denoted $\widehat{\eta}$
and is
called the Newcombian Class of $\eta$.

We naturally extend the target function $T_i$ to Newcombian Classes:
$T_i(\widehat\eta)=T_i(\eta)$, where $\eta$ is any element of
$\widehat\eta$.

If $T_i(\widehat\eta)=\emptyset$, $\widehat\eta$ is called a degenerate class.
\end{definition}

\begin{example}
In the assurance game, $T_2(o_1)=\{o_1\}$ and $T_2(o_1, n_2)=\{o_1\}$. This
means that they belong to the same equivalence class:
$\widehat{o_1}=\widehat{(o_1,n_2)}$. In particular, they can potentially discard
the exact same outcomes. Also, in the $\Gamma$-game, $T_2(n_2, n_1, n_2, n_1,
n_2)=T_2(n_2)$ since $(n_1, n_2, n_1, n_2)$ is degenerate. Hence $\widehat{(n_2,
n_1, n_2, n_1, n_2)}=\widehat{n_2}$.
\end{example}

Two target-equivalent Newcombian states
(i) have the same pure part, (ii) have an order like Newcombian
classes and (iii) are characterized with their worst payoff.

\begin{definition} (pure part of a non-degenerate Newcombian Class) Two
non-degenerate target-equivalent Newcombian States always have
the same pure part: had they not, then they could not target the
same outcomes. Consequently, we can define the pure part of
a non-degenerate Newcombian Class as being the pure part of any of
its elements: $\widehat\eta_1\equiv\eta_1$, for any $\eta \in
\widehat\eta$.
\end{definition}

\begin{definition}
(order of a class) We call order of a class the minimum of the orders of its
elements.
\end{definition}

\begin{example}
In the assurance game, the Newcombian Class $\widehat{o_1}\supset\{o_1, (o_1,
n_2)\}$ is of order $1$.
\end{example}

\begin{definition} \label{defworstpayoff} (worst payoff of a Newcombian Class)
Let $\widehat\eta$ be a non-degenerate Newcombian Class. The set of
$\widehat\eta$-targeted outcomes is finite and non-empty. Hence the
current player's payoff $\mathbf{payoff}_i = a_.^{p_{c_{i-1}}}$ has a minimum
value
on this set, which is $\mathbf{\widehat\eta}$'s worst
payoff:
\begin{equation}
\mathbf{wp}(\widehat\eta)=\min_{o\in
T_i(\widehat\eta)}\mathbf{payoff}_i(o).
\end{equation}
\end{definition}

\begin{proposition}
\label{propworstpayoff}
At any step $i$, a non-degenerate Newcombian Class is characterized
by its worst payoff.
\end{proposition}

\begin{example}
In the assurance game, at step 2, $\widehat{(n_2,
o_1)}$ is characterized by payoff 1.
\end{example}

\begin{remark}
As in Definition \ref{defworstpayoff}, we could define the best payoff
of a non-dege\-nerate class, but in the general case, we may have different
classes
with the same best payoff. The best payoff does not characterize a
class.
\end{remark}

Finally, the combination of Newcombian states (i.e., building a new, higher-order state 
with a node as the pure part and a Newcombian state as the discarding part) translates naturally to
Newcombian classes.

\begin{proposition}
\label{propclasseta}
Given a move $m\in F(c_{i-1})$ and a Newcombian Class $\hat{\eta}$, the
Newcombian Class of $(m,\eta)$ is the same for all representants $\eta$ of
$\hat{\eta}$. Consequently, one can define
$$(m,\hat{\eta})=\widehat{(m,\eta)}$$
for any $\eta \in \hat{\eta}$.
\end{proposition}

\begin{example}
In the assurance game, $(n_2, \widehat{o_1})=\widehat{(n_2, o_1)}$
(characterized by its worst outcome 1).
\end{example}

\subsubsection{The (explicit) current player's move}
\label{section-explicit-player-move}

We now give another formula to compute the next move, which is equivalent
to the one given in Definition
\ref{lemplayersmove} (the current player's move exists and is unique), with the
difference that here the proof gives its explicit construction (the
corresponding Newcombian State and the remaining, targeted outcomes). This is
necessary for the general algorithm to explicitly compute the Perfect Prediction
Equilibrium and also to determine the \emph{minimum level} used in the quick
algorithm (Section \ref{section-quick-algorithm}).

\begin{proposition}
\label{propuniqueeta}
Let $\widehat{\eta}^{(i)}$ be the only
class that discards any outcome targeted by the brothers of its pure
part ($\widehat{\eta}^{(i)}$ is also called best Newcombian Class\footnote{It is actually the maximum of a
a total order relation. A partial order relation can be defined on all
Newcombian States, and Newcombian Classes are defined so as to make this
relation a relation of total order on them.} at step $i$):

\begin{equation}
\forall n \in
F(c_{i-1})\setminus\{\textbf{pure}(\widehat\eta^{(i)})\}:\quad
T_i(n,\hat{\eta}^{(i)})=\emptyset
\end{equation}

This class $\widehat{\eta}^{(i)}$ exists and is unique.
\end{proposition}

\begin{example}
In the assurance game, $T_2(o_1, n_2, o_1)=\emptyset$ so that $\widehat{(n_2, o_1)}$ is
the class we are looking for.
In the $\Gamma$-game, $T_2(n_1, n_2, n_1, n_2)=\emptyset$ so that
$\widehat{(n_2, n_1, n_2)}$ is the class that discards any outcome in the other
subtrees than that of its pure part.
\end{example}

Now, if we consider any node $n \in
F(c_{i-1})\backslash\{\textbf{pure}(\widehat\eta^{(i)})\}$, i.e. any brother of
$\widehat\eta^{(i)}_1$, then every $n$-targeted outcome is worse
than the worst $\widehat\eta^{(i)}$-targeted outcome, in other words
$n$'s \emph{best payoff} is lower than $\widehat{\eta}^{(i)}$'s
\emph{worst payoff}.

At this stage, the second principle concludes that the current player moves on
to the pure
part of $\widehat{\eta}^{(i)}$. Later on, a player can
only consider outcomes in $T_i(\widehat{\eta}^{(i)})$, because had player
$p_{c_{i-1}}$ anticipated another outcome, he would have deviated.

\begin{proposition}
\label{propstructmove}
Then:
\begin{itemize}
\item the current player's move is $c_i = \textbf{pure}(\widehat{\eta}^{(i)})$
\item the remaining outcomes are $I_i=T_i(\widehat{\eta}^{(i)})$
\end{itemize}
\end{proposition}

\begin{example}
In the assurance game, $c_2=\mathbf{pure}(n_2, o_1)=n_2$ and $I_2$ = $T_2(n_2,
o_1)$ = $\{o_4\}$. In the $\Gamma$-game, $c_2=\mathbf{pure}(n_2, n_1, n_2)=n_2$ and
$I_2=T_2(n_2, n_1, n_2)=\{o_7, o_9, o_{10}, o_{11}\}$.

\end{example}

\subsubsection{An illustration of the constructions used in the proofs}

The proof of existence in Proposition \ref{propuniqueeta} gives
us an explicit algorithm to compute the current player's move.
The main idea is to recursively (over $k$) compute the Newcombian Class $\widehat\eta_{(k)}^{(i)}$ of order $k$
which has the highest worst payoff. At some point, one of these Newcombian
Classes $\widehat\eta^{(i)}$ is the one we are looking for.

The following example shows this construction.

\begin{example}
In game $\Gamma$, at step $2$ (remember that step 1 is directly defined, because
the game starts at the root), with respect to Peter (Fig.
\ref{Fig_Gamma_Target}):

There are two Newcombian Classes of order $1$: $\widehat{n_1}$ and
$\widehat{n_2}$. Their worst payoffs with
respect to Peter are, respectively, $-1$, and $0$.
$\widehat{n_2}$ has the highest worst payoff. Hence
$\widehat{\eta}_{(1)}^{(2)}=\widehat{n_2}$.

There is only one Newcombian Classes of order $2$: $\widehat{(n_1, n_2)}$.
$\widehat{(n_2, n_1)}$ is not of order $2$, because in fact it is equal to the
class
$\widehat{n_1}$. Hence we have
$\widehat{\eta}_{(2)}^{(2)}=\widehat{(n_1, n_2)}$.

We proceed by iteration:
\\
$\widehat{\eta}_{(1)}^{(2)}=\widehat{n_2}$ and the highest worst payoff is 1 (in
$n_1$'s subtree),
\\
$\widehat{\eta}_{(2)}^{(2)}=\widehat{(n_1, n_2)}$ and the highest worst payoff
is 2 (in $n_2$'s subtree),
\\
$\widehat{\eta}_{(3)}^{(2)}=\widehat{(n_2, n_1, n_2)}$,
\\
and finally since $T_1(n_1, n_2, n_1, n_2)=\emptyset$, we have
$\widehat{\eta}_{(3)}^{(2)}=\widehat{(n_2, n_1, n_2)}=\widehat\eta^{(2)}$.

\end{example}

Finally, the
proof of uniqueness in Proposition \ref{propuniqueeta} provides us with a
quick algorithm to determine the current player's move  - but by
skipping the procedure described in the proof of existence in Proposition
\ref{propuniqueeta}, we
cannot tell the order of the move, only its worst payoff and the
targeted outcomes.

\begin{figure}[htbp]
\centering
\includegraphics[width=8.4cm]{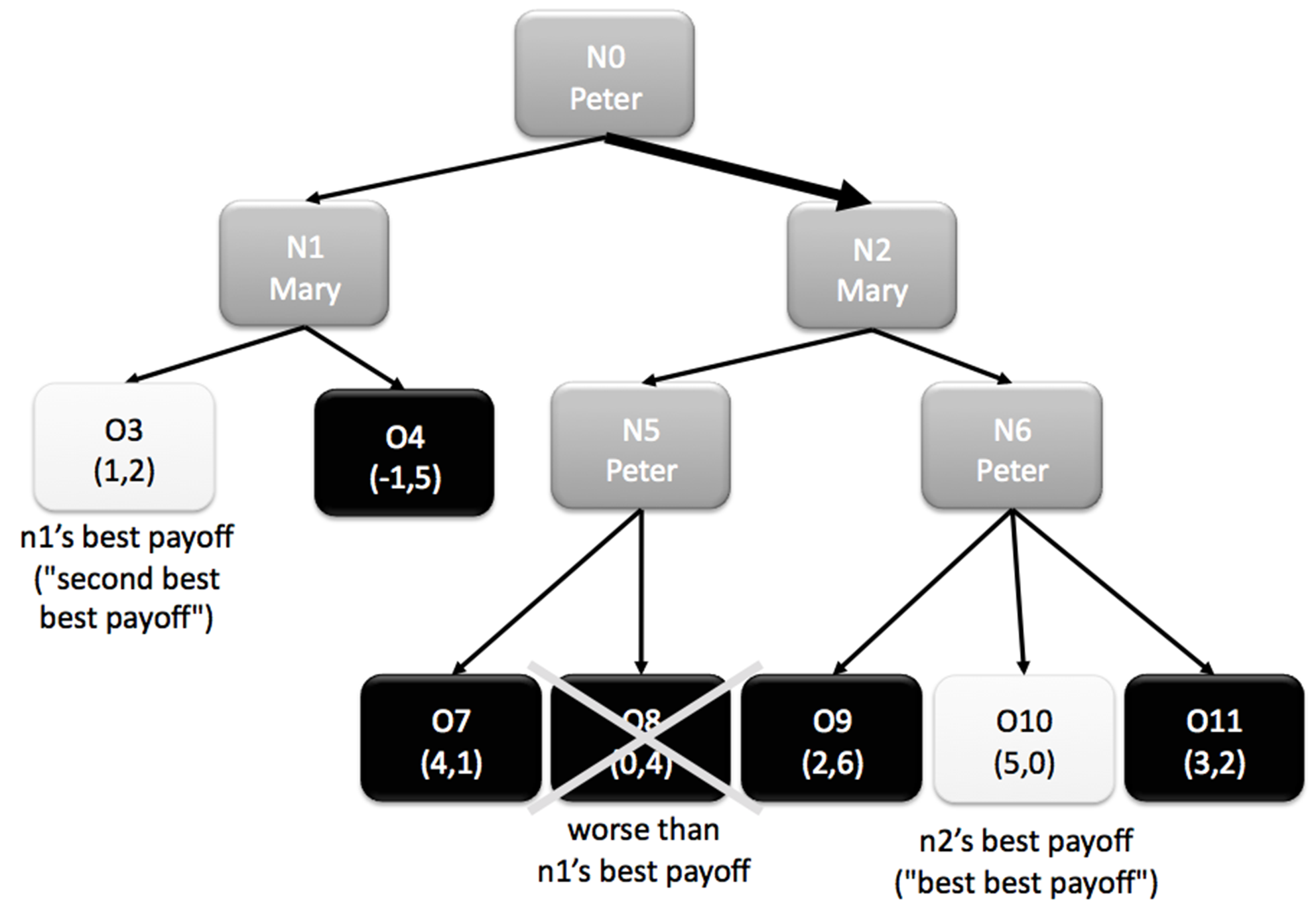}
\caption{Illustration of the quick algorithm to determine step $2$ in Game
$\Gamma$.
}\label{Fig_Gamma_Algo2}
\end{figure}

Therefore the preemption structure analysis indeed leads to the same results as
the original definition, and it is a powerful tool which provides us with two
algorithms to compute the equilibrium.

\subsection{The algorithms}

In this part the reader will find two algorithms computing the PPE as presented
in part 3 (see Fig. \ref{Fig_Gamma_Target} and \ref{Fig_Gamma_Algo2}):

- the general algorithm determines completely the Newcombian structure leading
to each \emph{current player's move}. It has been directly
derived from the proof of existence in Proposition \ref{propuniqueeta}, so that
this proof is a proof of correctness for the algorithm.

- the quick algorithm calculates directly the set of remaining outcomes at each
\emph{current player's move} by choosing the branch with the best outcome and
eliminating all outcomes below some minimum level. This algorithm has been
directly derived from the proof of uniqueness in Proposition
\ref{propuniqueeta}, so that this proof is also a proof of correctness for the
algorithm.

Both algorithms run in polynomial time with respect to the size of the tree.

\subsubsection{General algorithm}

This algorithm computes explicitly all intermediate discardings. It runs in
$O(n^3d)$ where $n$ is the number of nodes and leaves in the tree and $d$ its
depth.

\noindent The equilibrium path is called $(c_i)_{i=1..d}$.

\noindent The algorithm is initialized with:
\begin{center}
$c_1= n_0$ (the root)
\\
$I_1=\{$all outcomes of the game$\}$
\end{center}
Then a finite induction on $i$ gives all $(c_i)_{i=1..d}$ and all
$(I_i)_{i=1..d}$:

at each step $i$, a finite induction on $k$ gives a series of Newcombian States:$$
\{\widehat{\eta}_{(1)}^{(i)}\}\quad = \quad \arg\max _{
\begin{array}{c}
        \scriptstyle\widehat\eta \\
        \scriptstyle\mathrm{order} \left( \widehat\eta \right) = 1
      \end{array}}\min_{o\in T_i(\widehat\eta)}\qquad\mathbf{{payoff}}_i(o)
$$

$$
\{\widehat{\eta}_{(k+1)}^{(i)}\} \quad = \quad \arg\max _{
\begin{array}{c}
       \scriptstyle \widehat\eta \\
       \scriptstyle \widehat\eta=\widehat{(n,\eta_{(k)}^{(i)})} \\
       \scriptstyle n\in F(c_{i-1})\backslash\{\widehat\eta_{(k)1}^{(i)}\} \\
       \scriptstyle T_i(\widehat\eta)\neq\emptyset
      \end{array}}\min_{o\in T_i(\widehat\eta)}\qquad\mathbf{{payoff}_i(o)}
$$

and stops at some k, giving $\widehat{\eta}_{(k)}^{(i)} = \widehat{\eta}^{(i)}$

Then
\begin{itemize}
\item the current player's move is $c_i = \textbf{pure}(\widehat{\eta}^{(i)})$
\item the remaining outcomes are $I_i=T_i(\widehat{\eta}^{(i)})$
\end{itemize}

The induction ends when the path reaches an outcome: $$I_d=\{c_d\}$$

\subsubsection{Quick algorithm}

\label{section-quick-algorithm}

This algorithm calculates the equilibrium path without using the preemption
structure. It runs in $O(nd)$ where $n$ is the number of nodes and leaves in the
tree and $d$ its depth.

o $\leftarrow$ the root of the tree

While(o is not an outcome)

\hspace{20pt} p $\leftarrow$ current player at node o

\hspace{20pt} s $\leftarrow$ the subtree starting at an offspring of o

\hspace{44pt} with the maximum payoff for player p

\hspace{20pt} If(there are other subtrees than s)

\hspace{36pt} M $\leftarrow$ player p's maximum payoff in the other subtrees
than s

\hspace{36pt} Eliminate all other subtrees than s

\hspace{36pt} Discard in s all outcomes with a payoff lower than M for player p

\hspace{36pt} Clean the tree (remove all interrupted paths)

\hspace{20pt}  o $\leftarrow$ the root of s

Return o

\begin{example}
An example of application of this algorithm at the root is shown on Fig.
\ref{Fig_Gamma_Algo2}. Peter's highest payoff is at the $n_2$-targeted outcome
$o_{10}$: $5$.
Among the brothers of $n_2$, the subtree at $n_1$ has the highest best payoff:
$1$.
We discard all $n_2$-targeted outcomes whose payoff with respect
to Peter are lower than $1$: $o_8$. We get the same next move $c_2=n_2$ and the
same set $I_2=\{o_7, o_9, o_{10}, o_{11}\}$as by doing the full procedure at
step 2 (Fig.
\ref{Fig_Gamma_Target})
\end{example}

\section{Annex - Proofs of Lemmas, Theorems and Propositions}

\begin{proof} (Lemma \ref{lemplayersmove}) We prove each of the three points:

1. The outcome in $I_{i-1}$ with the highest payoff for the current player
$p_{c_{i-1}}$:
$$\quad \arg\max_{o\in I_{i-1}}a_o^{p_{c_{i-1}}}$$
cannot be discarded by a Newcombian State $\eta$. Its payoff would namely have
to be lower than the worst outcome of a non-empty subset $T_i(\eta)$ of
$I_{i-1}$, which is in contradiction with its definition.

2. Let us suppose that $I_i$ contains at least two outcomes descending from two
different offsprings of $c_{i-1}$.

Let $o_m$ be the worst outcome for $p_{c_{i-1}}$ in $I_i$ and $f_m$ its ancestor
among the offsprings of $c_{i-1}$. We will show that $o_m$ should have been
discarded and so cannot be in $I_i$.

By assumption, there is another offspring $f_p \ne f_m$ of $c_{i-1}$, with at
least one outcome which has not been discarded - let $o_p$ be the worst one of
those, and $(f_p, \eta)$ a Newcombian State. Because $o_p$ is not discarded,
$o_p \in T_i(f_p,\eta)$. By definition of $o_m$ and because of the strict
preference between two outcomes, $o_m$ is strictly worse than $o_p$ (which is
the worst outcome in $T_i(f_p,\eta)$), thus is preempted by $(f_p, \eta)$.

This is in contradiction with the fact that $o_m \in I_i$. $\Box$ 

\end{proof}

\begin{proof} (Theorem \ref{thmppe})
The induction presented in Lemma \ref{lemplayersmove} discards all outcomes but
one. Furthermore, no other reasoning can discard this last outcome. The two
players, having anticipated it, have no interest in modifying their choices and
will play towards this outcome. $\Box$
\end{proof}

\begin{proof} (Theorem \ref{thmpareto})

Suppose the outcome of the Perfect Prediction Equilibrium $o$ is not
Pareto-optimal. Then there must be some outcome $o^P$, which
Pareto-improves $o$:

$$a^{Peter}_o<a^{Peter}_{o^P}$$
$$a^{Mary}_o<a^{Mary}_{o^P}$$

(Both inequalities are strict because of the strict preferences).

This outcome $o^P$ has been discarded at some move $i$ by a Newcombian State
$\eta$ of order $k$.

One of two: 

- either $o \in T(\eta)$, in which case
the current player's payoff at $o$ is greater than (or equal
to) $\eta$'s worst payoff

- or $o\notin T(\eta)$, but since $o$ was not discarded at that stage, the
current player's payoff at $o$ is necessarily strictly greater than $\eta$'s
worst payoff

Since the current player's payoff at $o_p$ is lower than $\eta$'s worst payoff,
in both cases, we have contradicted the above inequalities. $\Box$
\end{proof}

\begin{proof} (Proposition \ref{propworstpayoff})
Given a payoff $x$, there is only one outcome $o$ that will give $x$
to the current player, in virtue of the assumption that all payoffs
are different. If two nodes $n$ and $n'$ in $F(c_{i-1})$ are different,
then they cannot target the same outcome: $n\neq n' \Rightarrow
T_i(n)\cap T_i(n')=\emptyset$. Hence there is a unique node $n$ in
$F(c_{i-1})$ such that outcome $o$ is the descendant of node $n$.

There is only one Newcombian Class whose pure part is $n$ and whose
worst payoff is $x$. This is because the Newcombian State $p$ discards all
outcomes worse than the worst
$p$-targeted outcome, whereas all better outcomes are not
discarded. Hence the required Newcombian Class is defined by the set
of $n$-targeted outcomes whose payoffs are greater than or equal to
$x$. $\Box$

\end{proof}

\begin{proof} (Proposition \ref{propclasseta})
Discarding only depends on the worst outcome of the discarding Newcombian State.
$\Box$
\end{proof}

\begin{proof} (Proposition \ref{propuniqueeta}) (Existence and uniqueness of
move $\widehat{\eta}^{(i)}$)

\emph{Uniqueness:} The pure part $n$ of the class $\widehat{\eta}^{(i)}$
necessarily targets the outcome with the highest payoff with respect
to the current player. If not, then $\widehat{\eta}^{(i)}$ could not
discard precisely this outcome with the highest payoff, which
contradicts the definition of $\widehat{\eta}^{(i)}$.

Among the brothers of $n$, we call $m$ the node which has the
highest best payoff. Then $\widehat{\eta}^{(i)}$'s worst payoff is
the minimum of the payoffs of the $n$-targeted outcomes, which are
(strictly) greater than $m$'s best payoff. A reason for it is that
it is the minimum level to discard all outcomes targeted by the
brothers of $n$. And it cannot be higher, since there is no other
class which could have discarded the outcome relative to this value.

According to Proposition \ref{propworstpayoff}, $\widehat{\eta}^{(i)}$ is
fully determined by its worst payoff.

\emph{Existence:} let us define a procedure based on several
\emph{little steps} $k$, which are to be distinguished from the \emph{step} $i$.
With respect to step $i$, we define the \textbf{best Newcombian Class
of order $k$} denoted $\widehat{\eta}_{(k)}^{(i)}$ by induction on little step
$k$.

Let us begin with $k=1$. The best Newcombian Class of order $1$,
$\widehat{\eta}_{(1)}^{(i)}$, is the Newcombian Class of order $1$ with the
highest worst payoff. This class exists, since 
Newcombian Classes of
order $1$ are identified with nodes in $F(c_{i-1})$, which are all
non-degenerate. Their worst payoffs are all
distinct (see Proposition \ref{propworstpayoff}), so
$\widehat{\eta}_{(1)}^{(i)}$ is unique.

\begin{equation}
\{\widehat{\eta}_{(1)}^{(i)}\}\quad = \quad \arg\max _{
\begin{array}{c}
        \scriptstyle \widehat\eta \\
        \scriptstyle \mathrm{order} \left( \widehat\eta \right) = 1
      \end{array}}\min_{o\in T_i(\widehat\eta)}\qquad\mathbf{{payoff}}_i(o)
\end{equation}

Suppose now we have defined $\widehat{\eta}_{(k)}^{(i)}$ for some
$k\geq1$. We have to distinguish between two cases:

1. Either $\widehat{\eta}_{(k)}^{(i)}$ discards any outcome targeted by
the brothers of its pure part $\textbf{pure}(\widehat\eta_{(k)}^{(i)})$, so
the procedure is
over and we have reached $\widehat{\eta}^{(i)}=\widehat{\eta}_{(k)}^{(i)}$.

2. Or there are still some brothers $n$ of
$\textbf{pure}(\widehat\eta_{(k)}^{(i)})$,
such that $(n,\widehat{\eta}_{(k)}^{(i)})$ is not degenerate. Then, by
definition, the best Newcombian class $\widehat{\eta}_{(k+1)}^{(i)}$ of
order $k+1$ will be the non-degenerate class
$(n,\widehat{\eta}_{(k)}^{(i)})$ with the highest worst payoff. As for
$\widehat{\eta}_{(1)}^{(i)}$, this class exists and is unique, because it
is the maximum of the injective function $\mathbf{wp}$ on a finite
set:

\begin{equation}
\{\widehat{\eta}_{(k+1)}^{(i)}\} \quad = \quad \arg\max _{
\begin{array}{c}
        \scriptstyle\widehat\eta \\
       \scriptstyle \widehat\eta=(n,\widehat{\eta}_{(k)}^{(i)}) \\
\scriptstyle n\in F(c_{i-1})\setminus\{\textbf{pure}(\widehat\eta_{(k)}^{(i)})\}
\\
       \scriptstyle T_i(\widehat\eta)\neq\emptyset
      \end{array}}\min_{o\in T_i(\widehat\eta)}\qquad\mathbf{{payoff}_i(o)}
\end{equation}

For each $n\in F(c_{i-1})\setminus\{\textbf{pure}(\widehat\eta_{(k)}^{(i)})\}$,
because
$(n,\widehat{\eta}_{(k)}^{(i)})$ is non-degenerate, there
are some $n$-targeted outcomes which have not been discarded by
$\widehat{\eta}_{(k)}^{(i)}$, i.e,
$\mathbf{wp}((n,\widehat{\eta}_{(k)}^{(i)}))>\mathbf{wp}(\widehat{\eta}_{(k)}^{(i)})$.
Since $\widehat{\eta}_{(k+1)}^{(i)}$ is one of these
$(n,\widehat{\eta}_{(k)}^{(i)})$,
$\mathbf{wp}(\widehat{\eta}_{(k+1)}^{(i)})>\mathbf{wp}(\widehat{\eta}_{(k)}^{(i)})$.

Since the set of outcomes is finite, the sequence
$(\mathbf{wp}(\widehat{\eta}_{(k)}^{(i)}))_{k\geq1}$ cannot increase
indefinitely: we will necessarily reach case $1$ for some finite
$k$. This shows the existence of $\widehat{\eta}^{(i)}$. $\Box$

\end{proof}

\begin{proof} (Proposition \ref{propstructmove}) (Equivalent definitions of
$c_i$ and $I_i$)

Let us denote $c_i$ and $I_i$ as the current player's move and the set of the
not-yet-discarded outcomes as defined in Lemma \ref{lemplayersmove}, and let us
denote $c'_i$ and $I'_i$ as they are suggested in Proposition
\ref{propstructmove}.

In the uniqueness proof of $\widehat{\eta}^{(i)}$ we mentioned that $c'_i$
necessarily targets the outcome $o$ with the highest payoff with respect to the
current player. This very outcome cannot be discarded, hence $o\in I_i$.
According to Lemma \ref{lemplayersmove}, all outcomes in $I_i$, thus $o$ as
well, are the descendants of $c_i$. Hence $c_i = c'_i$.

$I_i$ is defined as the set of all the outcomes that cannot be discarded at step
$i$.
First, we have $I_i \subset I'_i$ since a non-discarded outcome at step i must
be a descendant of $c_i$ (Lemma \ref{lemplayersmove}) and has to be targeted by
the best Newcombian Class (because it is not discarded).

Then, we have $I'_i \subset I_i$: all outcomes in $I'_i$ cannot be discarded by
anything, because the worst of them is strictly greater than the best outcome
among the descendants of the brothers of $c_i$ (Uniqueness proof of Proposition
\ref{propuniqueeta}).

Hence, $I_i=I'_i$ as well. $\Box$

\end{proof}

\section{Annex - Biped Games: complete analysis}

In this section, we perform a complete analysis of biped games and investigate
when the Subgame Perfect Equilibrium is the same as the Perfect Prediction
Equilibrium and when it is not.

\begin{figure}[htbp]
\begin{center}
\includegraphics[width=8.4cm]{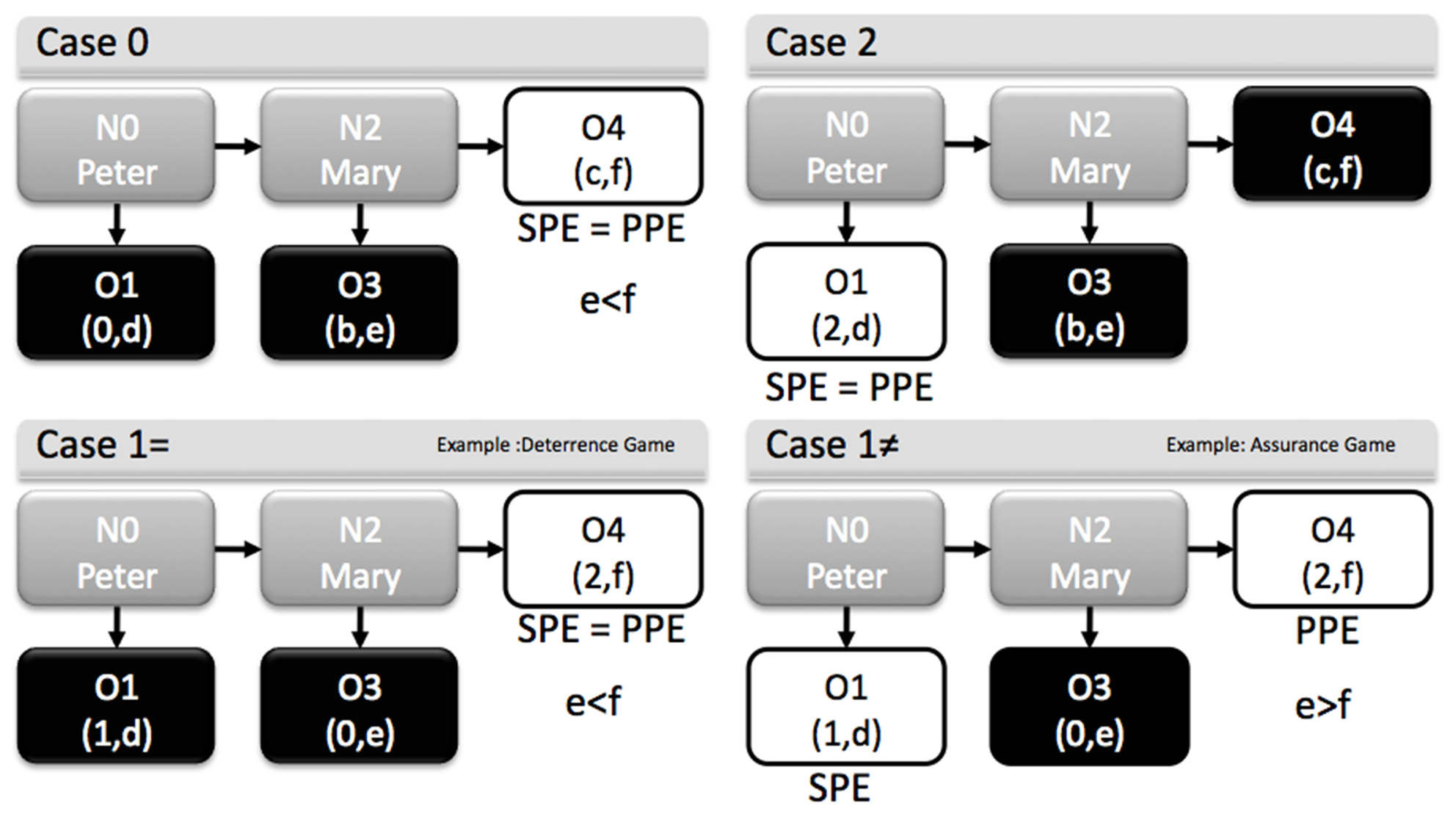}
\end{center}
\caption{Biped games: the four cases}\label{fourcases}
\end{figure}

In our framework, only the order relation (preferences) between the outcomes for
each player is relevant, so that without loss of generality we can assume that
the payoffs are 0, 1 and 2 for each player. If we label the payoffs $(a,d)$,
$(b,e)$ and $(c,f)$, we have $\frac{3!.3!}{2}=18$ different games. (We divide by
two because exchanging $(b,e)$ and $(c,f)$ does not change the game). Among
these 18 games, we can distinguish four cases.

\begin{itemize}
\item Case 0 ($a=0$, assuming $e<f$ without loss of generality - 6 different
games). Both equilibria reach outcome 4.
In the Subgame Perfect Equilibrium, if Mary plays at node 2, she will prefer
outcome 4 which gives her f, and Peter will get c. Since $c>0$ (no
indifference), Peter at node 0 will choose node 2.
In the Perfect Prediction Equilibrium, Peter cannot preempt anything but outcome
1 (with node 2, since any outcome stemming from node 2 is better than outcome 1)
so that Mary can choose her favorite outcome: outcome 4. This is the PPE.
\item Case 1 ($a=1$, assuming $b=0$, $c=2$ without loss of generality - 6
different games). This case splits into two subcases:
\begin{itemize}
\item Case 1= ($e<f$, i.e., Mary has the same preferences as Peter at node 2).
Both equilibria reach outcome 4.
In the Subgame Perfect Equilibrium, Mary playing at node 2 chooses outcome 4.
She gets f and Peter gets 2. Peter playing at node 0 will hence choose node 2.
In the Perfect Prediction Equilibrium, Peter preempts outcome 3 with outcome 0
and then plays node 2. Mary has no choice but playing outcome 4.
\item Case 1$\ne$ ($e>f$, i.e., Mary has opposite preferences to Peter at node
2). The equilibria are different.
In the Subgame Perfect Equilibrium, Mary playing at node 2 chooses outcome 2.
She gets e and Peter gets 0. Peter playing at node 0 will hence choose outcome
1, which is the SPE.
In the Perfect Prediction Equilibrium, Peter preempts outcome 3 with outcome 0
and then plays node 2. Mary has no choice but playing outcome 4, which is the
PPE.
\end{itemize}
\item Case 2 ($a=2$ - 6 different games). Both equilibria reach outcome 1.
In the Subgame Perfect Equilibrium, whatever Mary plays at node 2, Peter at node
0 will choose outcome 0 where he gets 2.
In the Perfect Prediction Equilibrium, Peter preempts with outcome 0 all
outcomes stemming from node 2 so that only outcome 0 remains. This is the PPE.
\end{itemize}

Hence, it is interesting to see that for 15 biped games out of 18, the Subgame
Perfect Equilibrium and the Perfect Prediction Equilibrium are the same. Only 3
games lead to different equilibria.

For one of these three games ($d=0$, $e=2$, $f=1$), the Subgame Perfect
Equilibrium encounters the Backward Induction Paradox (see Section \ref{section-backward-induction-paradox}), which is solved by
the Perfect Prediction Equilibrium. For the two other games ($d=2$, $e=1$, $f=0$ and $d=1$, $e=2$, $f=0$), Peter's preempting power is also stronger than Mary's propensity to deviate, leading though to a PPE whose payoff is worse for Mary than at the SPE ($o_1$).

\section{Annex - The PPE as the solution of a first-order-logic equation system}
\label{section-equations}

In this section, we present how a first-order-logic equation system can be built for each game, so that the PPE is the unique solution to this system. This sets up an unambiguous formal basis for defining the PPE.

The equations described in Section \ref{section-equations} induce a graph on the powerset of the outcomes: each equation
contributes an edge from a set of logically impossible outcomes to a bigger set of
logically impossible outcomes. Only the equations indexed on a vertex in the connected
component of the empty set (i.e., all sets of logically impossible outcomes that
can appear at all in any step of a reasoning) are considered for the system.
A reasoning can be seen as a walk in the powerset graph
induced by the first-order-logics equations, starting at the empty set. There
is a unique outcome which belongs to no set in the connected component of
the empty set, and this outcome is the outcome of the PPE.

\subsection{Framework}

The framework is the same as in this paper. $O$ is the set of all outcomes, $N$ is the set of all nodes, $F(n)$ is the set of the children of  node $n$. $D(n)$ is the set of the outcomes which are descendants of node $n$. $P(n)$ is the parent of node or outcome $n$. $p_n$ is the player playing at node $n$. $a_p^o$ is the payoff of player $p$ at outcome $o$.

\subsection{Reaction path of a tree (Second Principle)}

For each tree T, we say that a path R = $(r_i)_{0\le i\le l}$ starting at the root is a reaction path of $T$ if it fulfills the following conditions:

\begin{itemize}
\item $r_0$ is the root.
\item For any $0 < i \le l, r_{i} \in F(r_{i-1})$
\item $o \in D(r_{l})$
\item (Second principle) For any $0 < i \le l$: all outcomes in $D(r_i)$ are better than all outcomes descending from the siblings of $r_i$ for the player at $r_{i-1}$.
\item It is maximal (there is no longer reaction path).
\end{itemize}

A non-empty tree always has exactly one reaction path (possibly simply its root). This is denoted  $T \longrightarrow R$. Since a path is uniquely identified by its last node, we will also write $T \longrightarrow r_l$. Note that $r_l$ (the last node in R) is not necessarily an outcome.

\subsection{Preempting reaction path of a tree to an outcome (First Principle)}

For each tree T and each one of its outcomes $o$, we say that a path R = $(r_i)_i$ starting at the root is a preempting reaction path of $T$ to $o$ if it fulfills the following conditions:

\begin{itemize}
\item $(r_i)_{i<l}$ is a reaction path of $T$
\item $r_{l} \in F(r_{l-1})$
\item $o \notin D(r_l)$
\item (First principle) All outcomes in $D(r_l)$ are better than $o$ for the player at $r_{l-1}$.
\end{itemize}

This is denoted  $T \overset{o}\rightsquigarrow R$. Since a path is uniquely identified by its last node, we will also write $T \overset{o}\rightsquigarrow r_l$. Note that a preempting reaction path may not exist given a tree and an outcome.

\subsection{Outcome Powerset Graph}

More precisely, we can organize all sets of outcomes (subsets of $O$) in a graph as follows.

For each reaction path $T \setminus U \longrightarrow r$, there is an edge from $U$ to $U \bigcup (O \setminus D(r))$.

For each preempting reaction path $T \setminus U \overset{o}\rightsquigarrow r$, there is an edge from $U$ to $U \bigcup \{o\}$.

The semantics of an edge is: if all outcomes on the left cannot be in the solution, then all outcomes on the right cannot be either.

We call $\mathcal{U}$ the connected component of this graph containing the empty set. This corresponds to the sets of eliminated outcomes actually reachable in a reasoning.

\subsection{The system of equations}
\label{section-equations}
Given a tree T, the system of equation is built as follows. Its unknowns are the variables $S_n$ for every node or outcome $n$ in the tree. $S_n$ is true if and only if $n$ is on the equilibrium path.

The PPE is defined as the (unique) solution of the system of equations

\begin{equation*}((C_n)_{n \in (N \cup O)}, (P^1_e)_{e \in E_1}, (P^2_e)_{e \in E_2})\end{equation*}

as follows.

\subsubsection{The causal bridge equations}

First, the equations maintaining the causal bridge are the following, for any node or outcome $n$:
\begin{equation*}
S_n \Rightarrow S_{P(n)} \wedge \bigwedge_{s \in F(P(n)) \setminus \{n\}} \bar{S_s}
\tag{$C_n$}
\end{equation*}

i.e., if a node or outcome is in the solution, then its parent should be as well, and none of its siblings should be. This forces the solution to be a path on the tree.

\subsubsection{The second-principle equations}

Secondly, the second-principle equations are indexed on indices (U, r) in the subset $E_2$ of $\mathcal{U} \times O$ corresponding to reaction paths $T \setminus U \longrightarrow r$:

\begin{equation*}
\bigwedge_{p \in U} \bar{S_p} \Rightarrow S_r
\tag{$P^2_{U, R}$}
\end{equation*}

\subsubsection{The first-principle (preempting) equations}

Finally, the first-principle equations are indexed on indices (U, o, r) in the subset $E_1$ of $\mathcal{U} \times O \times O$ corresponding to preempting reaction paths $T \setminus U \overset{o}\rightsquigarrow r$:

\begin{equation*}
(\bigwedge_{p \in U} \bar{S_p}) \wedge  S_{o} \Rightarrow S_r
\tag{$P^1_{U,o, R}$}
\end{equation*}

\subsection{Example for the assurance game}

There are five  variables here: $S_o = [\text{root} \in S], S_1 = [(P,D) \in S]$, $S_2 = [(P,C) \in S]$, $S_3 = [(M,D) \in S]$, $S_4 = [(M,C) \in S]$ (here we use edge labels to allow comparison with Dupuy's original equations).

The reaction paths (and the corresponding edges in the outcome powerset graph) are:

\definecolor{selected}{rgb}{0.7, 0.2, 0.2}

\begin{equation}
\begin{array}{rcl}
{\color{selected} T \longrightarrow n_0} & {\color{selected} \text{ contributing edge }} & {{\color{selected} \emptyset \longrightarrow \emptyset}}\\
T \setminus \{o_1\} \longrightarrow o_3  & \text{ contributing edge } & \{o_1\} \longrightarrow \{o_1, o_4\}\\
{\color{selected} T \setminus \{o_3\} \longrightarrow o_4  }& {\color{selected} \text{ contributing edge }} & {\color{selected} \{o_3\} \longrightarrow \{o_1, o_3\}}\\
T \setminus \{o_4\} \longrightarrow o_1  & \text{ contributing edge } & \{o_4\} \longrightarrow \{o_3, o_4\}\\
{\color{selected} T \setminus \{o_1, o_3\} \longrightarrow o_4 }& {\color{selected} \text{ contributing edge }} & {\color{selected} \{o_1, o_3\} \longrightarrow \{o_1, o_3\}} \\
T \setminus \{o_1, o_4\} \longrightarrow o_3  & \text{ contributing edge } & \{o_3, o_4\} \longrightarrow \{o_3, o_4\} \\
T \setminus \{o_3, o_4\} \longrightarrow o_1  & \text{ contributing edge } & \{o_3, o_4\} \longrightarrow \{o_3, o_4\} \\
\end{array}
\end{equation}

In this example, only one preempting reaction path can be found:
\begin{equation}
\begin{array}{rcl}
{\color{selected} T \overset{o_3}\rightsquigarrow o_1} & {\color{selected} \text{ contributing edge } }& {\color{selected} \emptyset \longrightarrow \{o_3\}}\\
\end{array}
\end{equation}

The edges in the connected component of the empty set are marked in red.

\begin{figure}[htbp]
\begin{center}
\includegraphics[width=4cm]{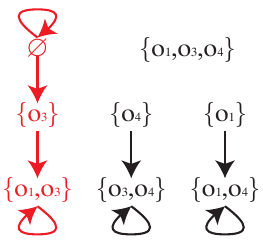}
\end{center}
\caption{The Outcome Powerset Graph for the promise game}
\label{Fig_PowersetGraph}
\end{figure}

Hence, the equations are:

\begin{align*}
\text{Causal bridge} \\
S_0 \Rightarrow \text{true} \tag{$0: C_0$} \\
S_1 \Rightarrow S_0 \wedge \bar{S_2} \tag{$1: C_1$} \\
S_2 \Rightarrow S_0 \wedge \bar{S_1} \tag{$2: C_2$} \\
S_3 \Rightarrow S_2 \wedge \bar{S_4} \tag{$3: C_3$} \\
S_4 \Rightarrow S_2 \wedge \bar{S_3} \tag{$4: C_4$} \\
\text{First Principle} \\
S_3 \Rightarrow S_1 \tag{$5: P^1_{\emptyset, o_3, o_1)}$} \\
\text{Second Principle} \\
\text{true}\Rightarrow S_0 \tag{$6: P^2_{\emptyset, n_0)}$} \\
\bar{S_3} \Rightarrow S_4 \tag{$7: P^2_{\{o_3\}, o_4)}$} \\
\bar{S_1} \wedge \bar{S_3} \Rightarrow S_4 \tag{$8: P^2_{\{o_1, o_3\}, o_4)}$} \\
\end{align*}

Dupuy's equations correspond to (5) and (7), which imply $S_0, \bar{S_1}, S_2, \bar{S_3}, S_4$, i.e.,  $\text{root} \in S$, $(P,D) \notin S$, $(P,C) \in S$, $(M,D) \notin S$ and $(M,C) \in S$ (this is our equilibrium).

Let us show this again.

From (2) and (3), one can deduce $S_3 \Rightarrow \bar{S_1}$, so that because of (5),  $\bar{S_3}$.

From (7) one deduces $S_4$, from (4) $S_2$ and from (2) or (6), $S_0$.

Is this a solution? With this assignment, the left part of (1), (3), (5) is false, making these implications all true, and in (0), (2), (4), (6), (7), (8), the left part is true, and the right part also, making the implications true as well.

Hence, it is proven that $S_0, \bar{S_1}, S_2, \bar{S_3}, S_4$ is the one and only solution of the system.

\section{Annex - Link with Newcombian states}

At step $i$ of the reasoning presented in Section 3, given the eliminated outcomes $I_{i-1}$, building a Newcombian state $(\eta, p)$ corresponds to the following
reasoning (we abuse notations by using a node for its number).

For the first principle (preempting an outcome $o$), using equation $$P^1_{(O \setminus (I_{i-1}) \cup P_i(\text{discard}(p)), o, \eta}$$
and equation $C_o$ leads for $S_o$ to be false.

For the second principle (choosing $\eta$), using equation $$P^2_{(O \setminus (I_{i-1}) \cup P_i(\text{discard}(p)), \eta}$$ and equation  $C_{\eta}$ leads for $S_{\eta}$ to be true and for $S_o$ to be false for each sibling $o$ of $\eta$ .

\section{Annex - The special case of invertible trees}

The assurance game and Take-or-Leave game are special cases of games, which
are called invertible trees in \citet{JPDPFNCE}. They were the starting point
for defining the PPE. Their trees
detain a natural chronology: at each node, the outcome below the node is in the
present, whereas the previous node on the left is in the past and the next one
on the right is in the future.

For these games, the procedure is rather simple and we can give a simpler
version of the algorithm (but it derives from the same procedure as in the
general construction). At each node, starting from the root, the current player
compares the present outcome with the outcomes in the future:

1. if the present outcome is better than all outcomes in the future, the
procedure stops: the PPE outcome is the present one (the future outcomes are all
preempted by the present Take-move)

2. if all outcomes in the future are better than the present one, then the
player Leaves (the present outcome is preempted by the Leave-move)

3. if some outcomes in the future are worse and some better than the present
outcome, then first all worse outcomes in the future are preempted by the
present Take-move, and then the present outcome is discarded like in point 2.

\end{document}